\documentclass[11pt]{article}

\usepackage{hyperref}
 
\pdfoutput=1
 
\usepackage[english]{babel}
 
\usepackage{graphicx}
\usepackage{arydshln}
\usepackage{lineno}
\usepackage{wrapfig}
\usepackage{subfigure}
\usepackage{rotating}
\usepackage{amssymb}
\usepackage{mathptmx}
\usepackage{scrextend}
\usepackage{multirow}
\usepackage{cite}

\usepackage[T1]{fontenc}
\usepackage[utf8]{inputenc}
\usepackage{babel}
\usepackage[font=small,labelfont=bf]{caption}

\usepackage[numbers]{natbib}

\usepackage[export]{adjustbox}

\begin{document}

\title{Characterization of the Hamamatsu \itshape R11265-103-M64 \upshape multi-anode photomultiplier tube}

\author{L. Cadamuro$^{b}$, M.Calvi$^{ab}$, L. Cassina$^{ab}$\thanks{Corresponding author. E-mail: lorenzo.cassina@mib.infn.it}~, A. Giachero$^{ab}$, C. Gotti$^{ab}$, 
\\B. Khanji$^{a}$, M. Maino$^{ab}$\thanks{Corresponding author. E-mail: matteo.maino@mib.infn.it}~, C. Matteuzzi$^{a}$ and G. Pessina$^{ab}$
\\
\\
\llap{$^a$} INFN, Sezione di Milano Bicocca, 
\\Piazza della Scienza 3, 20126, Milano, Italy\\
\llap{$^b$} Dipartimento di Fisica G. Occhialini, Universit\`a degli Studi
\\di Milano Bicocca, Piazza della Scienza 3, 20126, Milano, Italy\\}

\maketitle

\begin{abstract}

The aim of this paper is to fully characterize the new multi-anode photomultiplier tube \itshape R11265-103-M64\upshape, produced by Hamamatsu. Its high effective active area (77\%), its pixel size, the low dark signal rate and the capability to detect single photon signals make this tube suitable for an application in high energy physics, such as for RICH detectors. Four tubes and two different bias voltage dividers have been tested. The results of a standard characterization of the gain and the anode uniformity, the dark signal rate, the cross-talk and the device behaviour as a function of temperature have been studied. The behaviour of the tube is studied in a longitudinal magnetic field up to 100 Gauss. Shields made of a high permeability material are also investigated. The deterioration of the device performance due to long time operation at intense light exposure is studied. A quantitative analysis of the variation of the gain and the dark signals rate due to the aging is described.


\end{abstract}

\section{Introduction}

Recently Hamamatsu started to produce the new \hbox{\itshape R11265-103-M64\upshape} multi-anode photomultiplier tube. It is a 64-channel (8 x 8) pixel device with an active area of \hbox{23 x 23 mm$^{2}$} and a pixel size of approximately 2.9 x 2.9 mm$^{2}$ which is able to detect single photons. A very small inactive border around the device ensures a total active area coverage of about 77\%, while the MaPMT square cross-sectional geometry allows for a close packing ratio (approximately 90\%). These features make the \itshape R11265-103-M64 \upshape particularly tailored for an application in high energy physics, such as in a Ring Imaging Cherenkov detectors (RICH) which are used for the particles identification. In this case, the MaPMTs are supposed to add a minimal noise contribution with respect to the signal rate and a negligible cross-talk between neighbouring pixels. In addition to that, the photodetectors must not be affected by the magnetic field or, otherwise, external magnetic shields have to be developed. Considering, for instance, the RICH detector \cite{bib1} of LHCb \cite{bib2}, a magnetic field up to 30 G (3 mT) could be present on the photosensitive plane. Note that, although hybrid photon detectors (HPDs) are currently used in the RICH system of LHCb, MaPMT devices coupled with a new read-out electronics are foreseen to be employed for the LHCb RICH upgrade \cite{bib3}. This will allow the detector to run at a luminosity up to ten times higher with respect to the current value and calls for a higher read-out speed from all the subdetectors. Two different tubes made by Hamamatsu have already been characterized to verify their compliance with the LHCb RICH requirements: 
\begin{enumerate}
\item the \itshape H9500\upshape was rejected because of a non-negligible cross-talk between pixels and was not designed for single photon signal \cite{bib5}; 
\item the \itshape R7600-03-M64 \upshape which gives good performance for single photon detection \cite{bib6}, but has poor coverage. 
\end{enumerate}
The new \itshape R11265-103-M64 \upshape MaPMT is similar to the \itshape R7600-03-M64 \upshape but has an improved sensitive area. The measurements performed on the first available samples are presented in this paper. In order to fully characterize the device, different features were considered;  for each of them, the description of the setup conditions and the results obtained are described in this paper. In section \ref{sec:GeneralFeatures} the studies of the gain variation, anode uniformity, dark current, bias voltage divider, cross-talk and temperature bahaviour are presented. In section \ref{sec:Environment} the measurements of the device inside a magnetic field and aging monitoring are described.


\section{General features} \label{sec:GeneralFeatures}

The setup used to measure the basic characteristics of the tubes is the same as described in \cite{bib6}. All the MaPMT pixels, except those under study, were covered with a black mask and black tape. A commercial blue LED was biased with a very small current so that only few photons per second were generated. An optical fiber was positioned laterally to the LED and a small fraction of photons were brought to a pixel of the MaPMT, so that only that pixel was illuminated in each measurement. The signals at the anodes were read-out with a commercial wide bandwidth current feedback operational amplifier (CFOA) \cite{bib16}. The waveforms were then acquired and recorded with a Tektronics DPO7254 fast oscilloscope.


\subsection{Gain Variation and Anode Uniformity}

\begin{figure}[h!]
	\centering
		\begin{minipage}[t]{.48\textwidth}
			\includegraphics[width=1\textwidth]{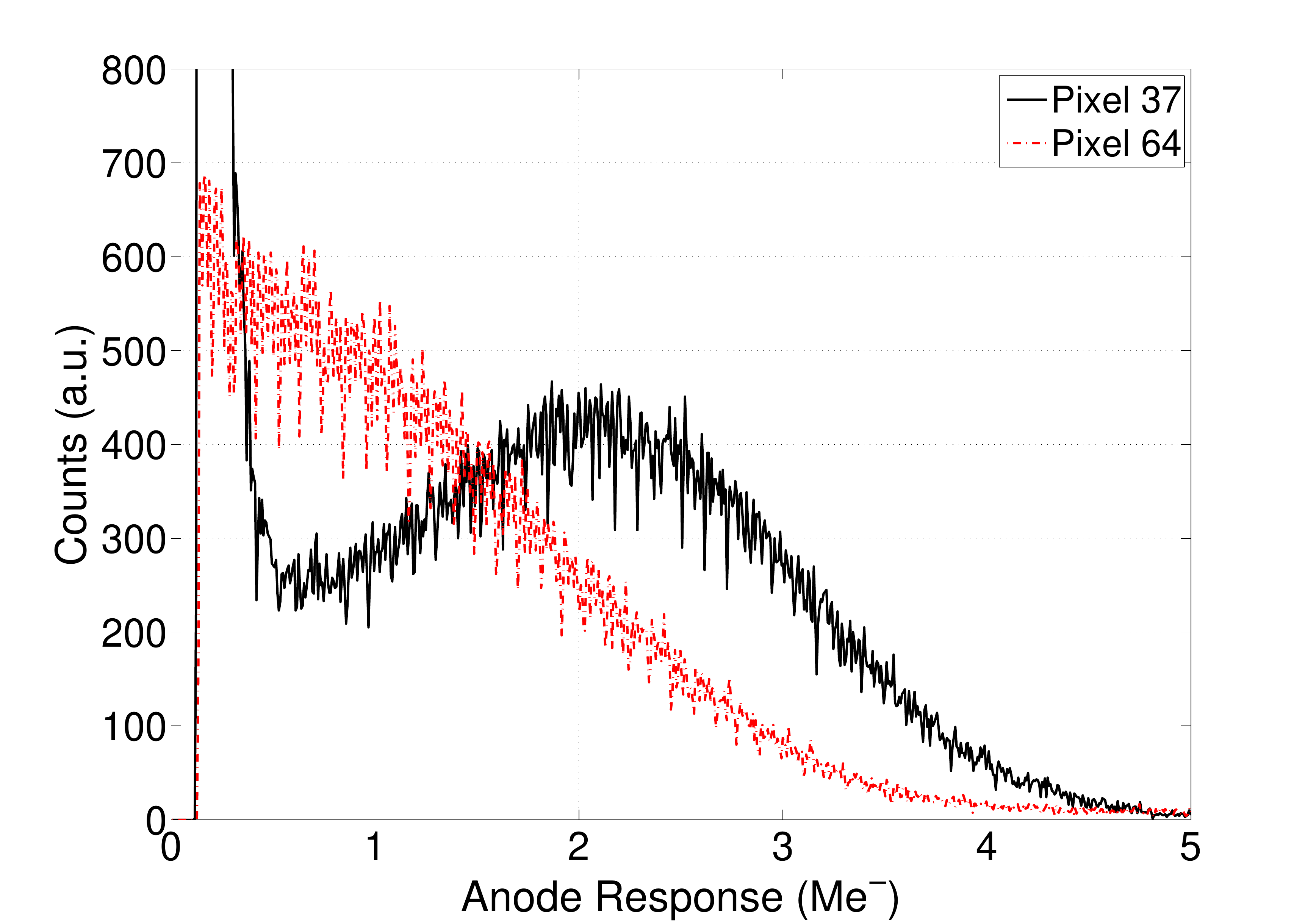}
			\caption{Single photon spectra for pixel 37 and 64 (MaPMT SN-ZN0702 biased at -1050 V.)}
			\label{fig:response}
		\end{minipage}%
	\hspace{5mm}%
		\begin{minipage}[t]{.48\textwidth}
			\includegraphics[width=1\textwidth]{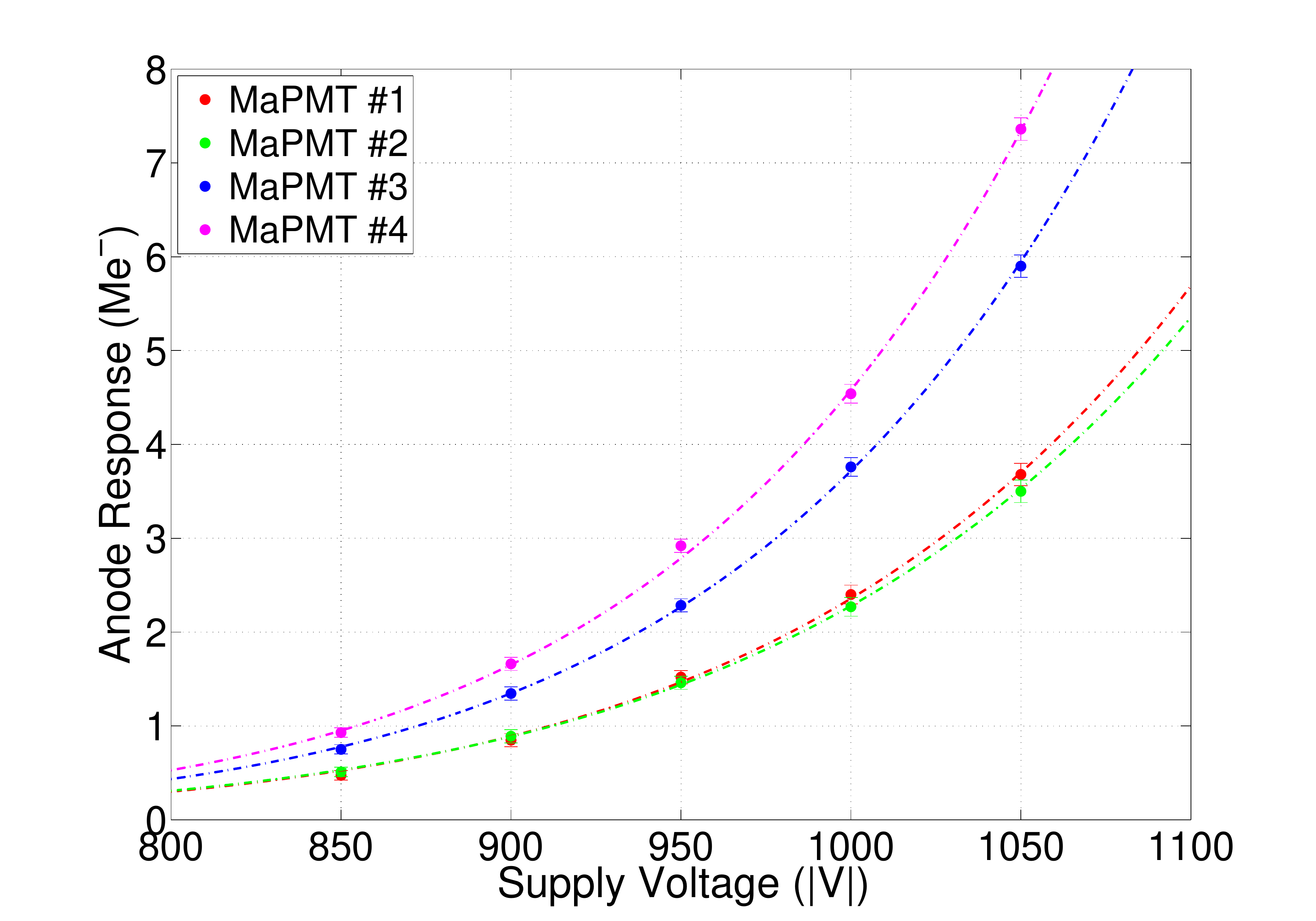}
			\caption{The single photon response versus the MaPMT bias voltage for the pixel with the maximum gain in each tube.}
			\label{fig:gain}
		\end{minipage}
\end{figure}

\begin{figure}[h!]
	\centering
	\begin{minipage}[t]{.48\textwidth}
	\centering
		\includegraphics[width=.67\textwidth]{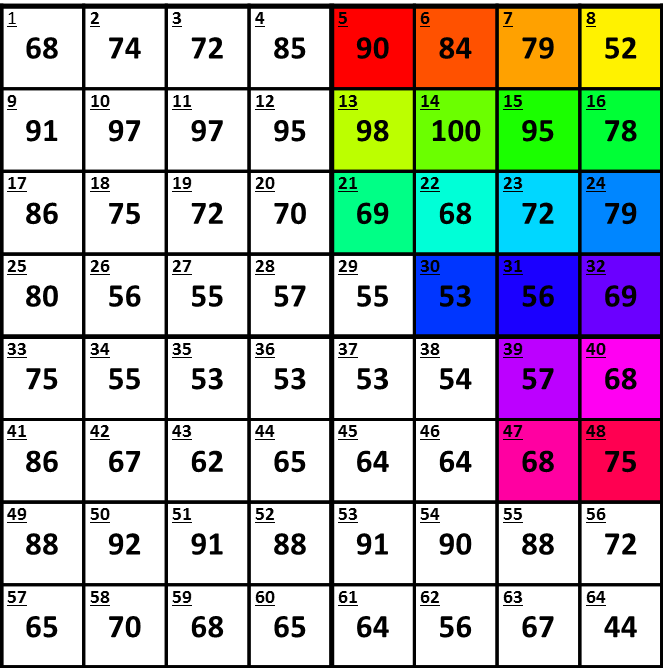}
		\end{minipage}%
	\hspace{5mm}%
		\begin{minipage}[t]{.48\textwidth}
		\centering
		\includegraphics[width=1\textwidth]{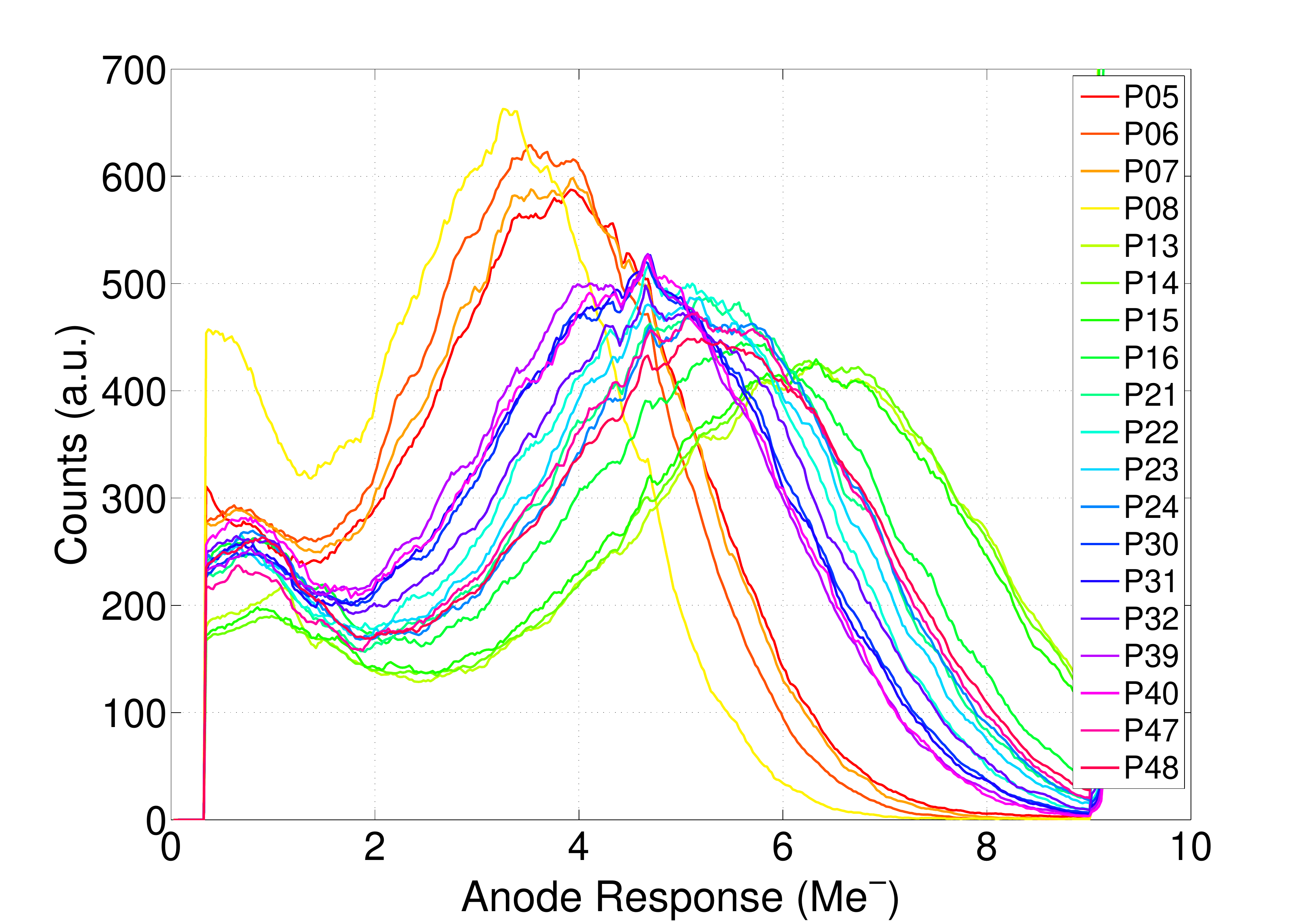}
		\end{minipage}%
		\caption{Pixel-to-pixel gain spread relative to the MaPMT SN-ZN0170. On the left, the uniformity table provided by Hamamatsu. On the right, the single photon spectra for some pixels highlighted with the corresponding colors (bias voltage equal to -950 V).}
	\label{fig:uniformity1}
\end{figure}

The MaPMT gain and the anode uniformity were measured at different bias voltages in order to compare them with the Hamamatsu datasheet. During these tests the standard bias divider with a 2.3-1.2-1-...-1-0.5 ratio from the first dynode to the last was used. Excellent single-photon response is observed in almost all channels and only few channels show a bad single-photon peak resolution (fig.\ref{fig:response}).
A MaPMT-to-MaPMT (fig.\ref{fig:gain}) and channel-to-channel variation in gain is measured up to a factor of three. In order to equalize the response, the channels to channel mismatch can be corrected by adjusting the MaPMT voltage supply and the gain or the threshold of the front-end electronics \cite{bib99}. A uniformity table is provided by Hamamatsu for every tube. For one of the devices tested at -900 V it is shown in fig.\ref{fig:uniformity1} (left). The numbers represent the relative gain of each pixel, the highest being 100. 

This measurement was performed by Hamamatsu with a continuous light source and thus the numbers reported represent the contributions from quantum efficiency, collection efficiency and gain. Figure \ref{fig:uniformity1} (right) shows the comparison of the spectra of different pixels, obtained with the setup previously described. The spectra are obtained with single photon signals and are normalized to the number of entries. The gain spread between different pixels is clearly visible. Almost all the pixels match the uniformity table, except for few of them (for example pixel 5). The gain spread among pixels is about 2.3 to 1 for this device. 

\subsection{Dark current}

\begin{figure}[h!]
	\centering
			\includegraphics[width=0.48\textwidth]{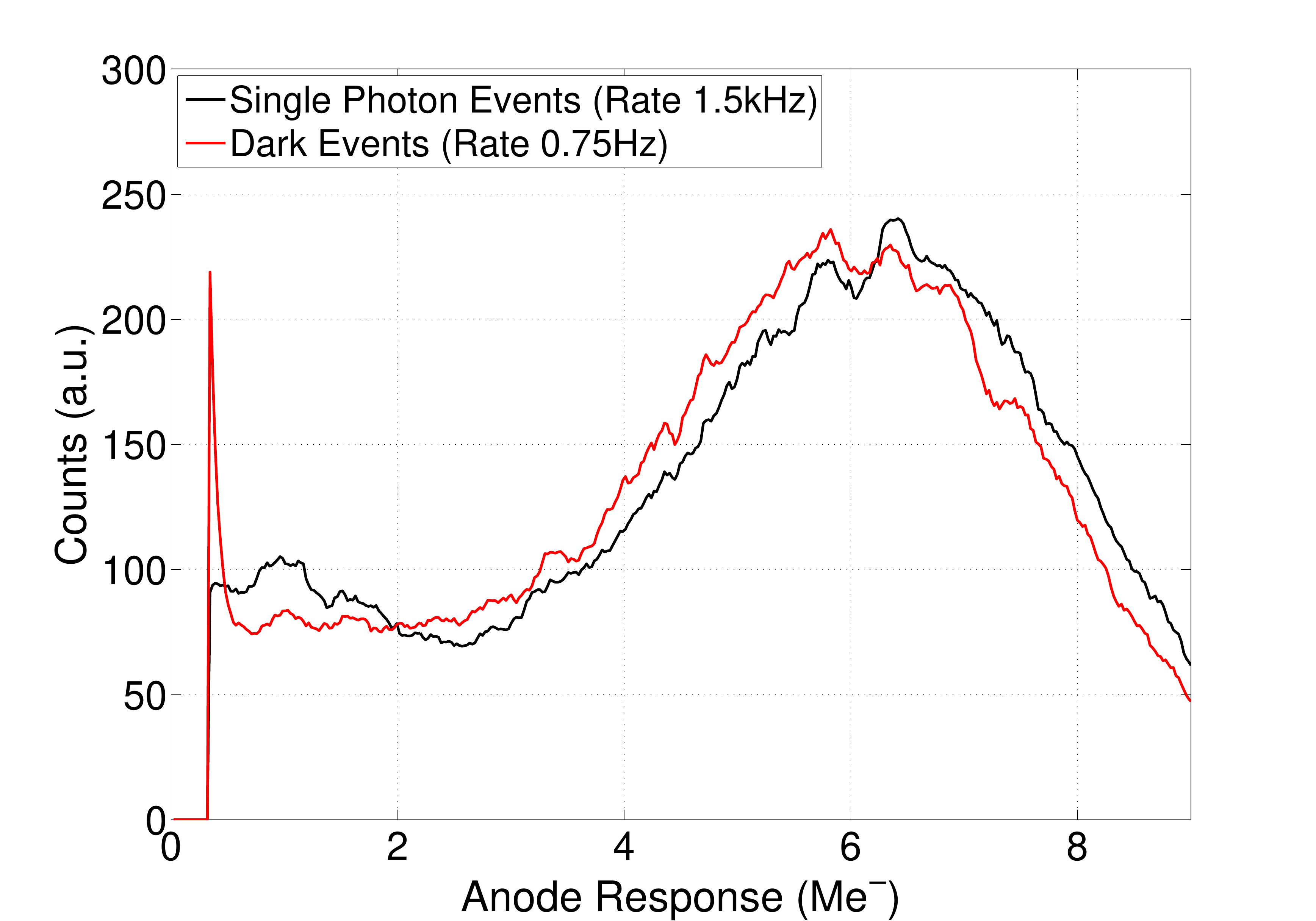}
			\caption{Comparison between the single photon spectra for dark and single photon events (MaPMT \hbox{SN-ZN0170} biased at -950V).}
			\label{fig:dark}
\end{figure}

The photodector dark current is measured at room temperature by recording the events above a threshold of 300 ke$^{-}$. Four MaPMTs have been tested in similar conditions and they showed the same behaviour. The dark signal rate is below 5 Hz per pixel, or about \hbox{60 Hz/cm$^{2}$}, for almost all anodes. Only few channels showed a slightly higher rate. The distribution of dark signals is acquired and compared with the spectra obtained turning on the LED. Figure \ref{fig:dark} shows the two spectra, normalized to the same number of counts in the single photon region: the spectra are very similar.

\subsection{Bias Divider}\label{sectiondivider}

\begin{figure}[h!]
  \begin{minipage}[b]{0.48\textwidth}
	\centering
		\scriptsize
			\begin{tabular}{c||c|c||c|c}
			\multicolumn{1}{c||}{Bias Divider} & \multicolumn{2}{c||}{Standard} & \multicolumn{2}{c}{Customized}\\ [0.5ex]
			\hline
			Stage & Ratio & $|\Delta V|$ & Ratio & $|\Delta V|$\\ [0.5ex]
			\hline
			K - P& 15 & 1000 & 18.4 & 1000\\
			\hline
			K - Dy1& 2.3 & 153 & 2.8 & 152\\
			Dy1 - Dy2& 1.2 & 80 & 1.5 & 81\\
			Dy2 - Dy3 & 1 & 67 & 1 & 54\\
			Dy3 - Dy4 & 1 & 67 & 1 & 54\\
			Dy4 ... Dy9 & 1 & 67 & 1 & 54\\
			Dy9 - Dy10 & 1 & 67 & 1 & 54\\
			Dy10 - Dy11 & 1 & 67 & 1 & 54\\
			Dy11 - Dy12 & 1 & 67 & 1.3 & 71\\
			Dy12 - G.R. & 1 & 67 & 3.3 & 179\\
			G.R. - P & 0.5 & 33.5 & 0.5 & 27\\
			\hline
			\end{tabular} 
		\captionof{table}{Table of the high voltage distribution ratio for the standard and customized bias divider (designed by Hamamatsu).}
		\label{table:divider1}
    \end{minipage}
  \hspace{5mm}%
  \begin{minipage}[b]{0.48\textwidth}
    \centering
    \includegraphics[width=1\textwidth]{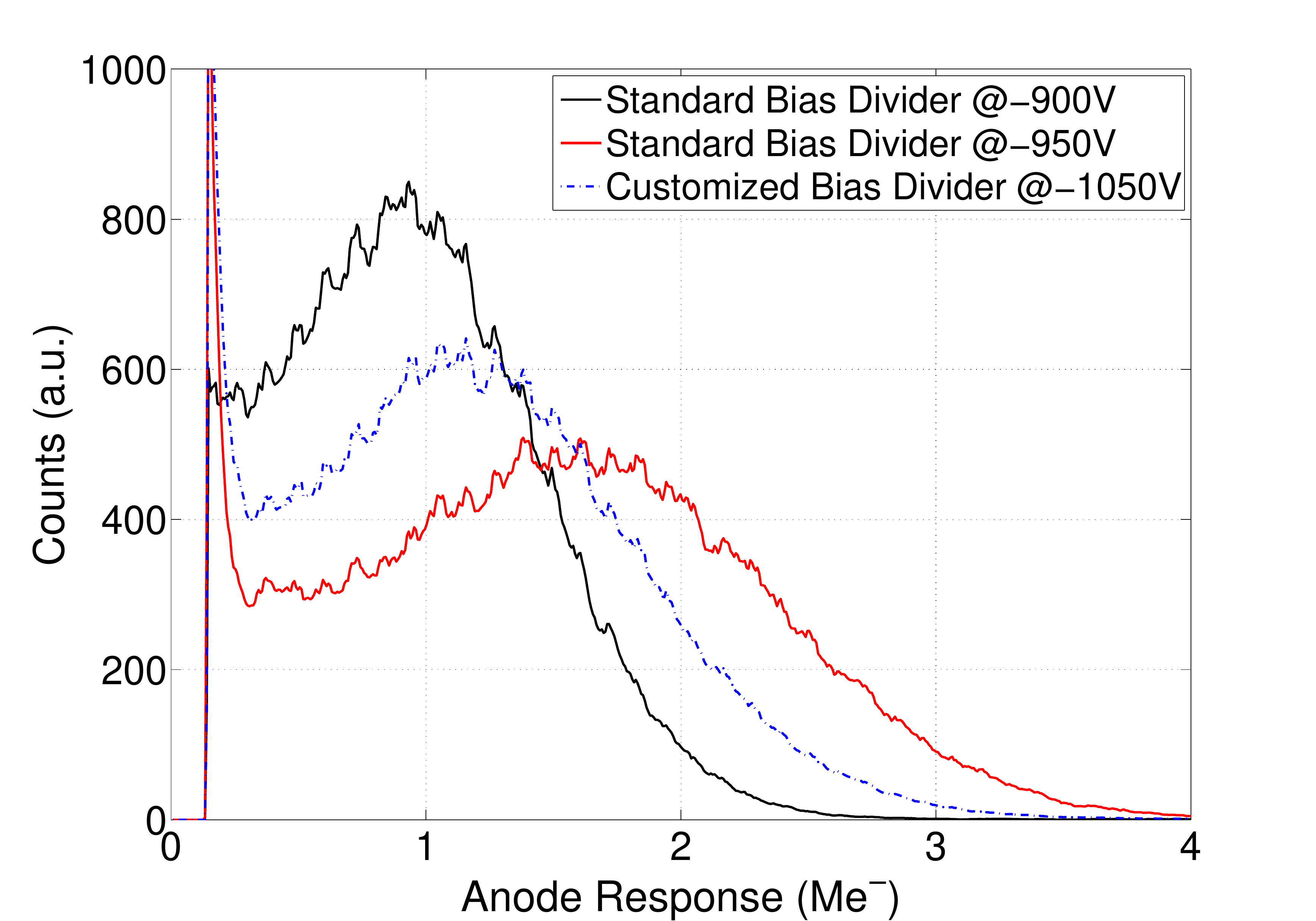}
    \captionof{figure}{Comparison between the single photon spectra acquired using the two different voltage dividers (MaPMT SN-ZN0170).}
    \label{fig:divider}
  \end{minipage}
 \end{figure}

The standard bias divider proposed by Hamamatsu splits the high voltage between dynodes with a 2.3-1.2-1-...-1-1-1-0.5 ratio. This configuration allows to obtain a mean gain of about 2 Me$^{-}$ at -1000 V supply voltage (fig.\ref{fig:divider}) and ensures a good linearity. From the device specifications, the maximum anode output current should be mantained below 100 $\mu$A. In the region of the LHCb RICH with a high occupancy, this limit value could be easily exceeded unless operating with a gain lower than 1 Me$^-$. If this is obtained only by lowering the high voltage, also the collection efficiency may decrease. For this reason we have tested a different bias divider \hbox{(2.8-1.5-1-...-1-1.3-3.3-0.5)} that provides an overall gain lower than the standard bias divider keeping the voltage unchanged at the first stages, as summarized in table \ref{table:divider1}. In this way the detection efficiency and in particular the photoelectrons collection efficiency is not affected. Note that this second bias divider has been also designed to increase the voltage at the last stage in order to maximize the electron collection just before the anode pins. Both the dividers have not been optimized independently, but they have been suggested by Hamamatsu. For all the following tests presented in this paper, the standard bias divider was used.

\subsection{Cross-talk}

\begin{figure}[htbp]
	\centering
		\begin{minipage}[b]{.4\textwidth}
			\includegraphics[width=1\textwidth]{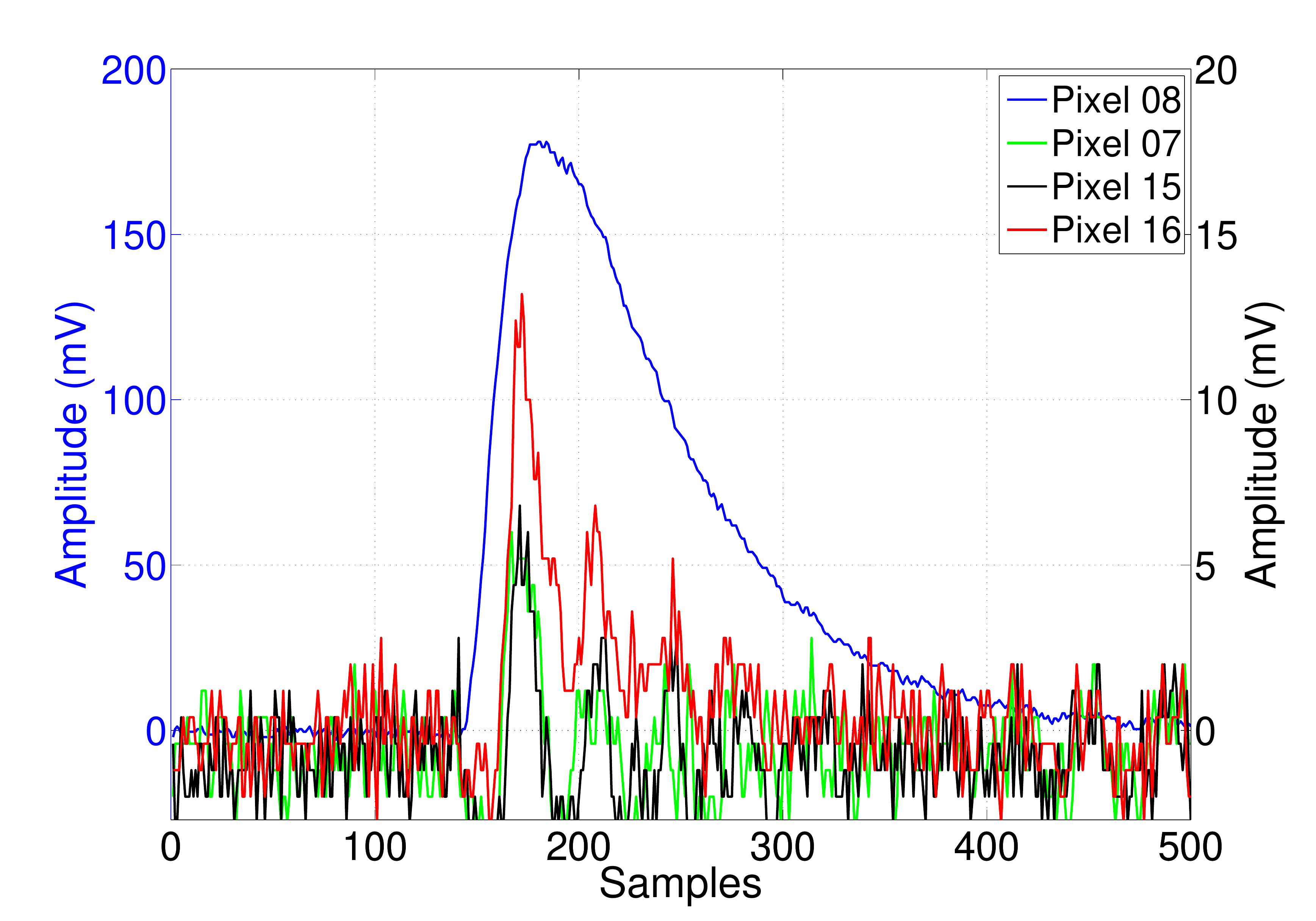}
		\caption{A single photon event (left y-axis) and in coincidence the neighbouring pixel events (right \hbox{y-axis}), magnified by a factor of 10.}
			\label{fig:cross-talk}
		\end{minipage}%
	\hspace{3mm}%
		\begin{minipage}[b]{.565\textwidth}
		\centering
		\includegraphics[width=0.48\textwidth]{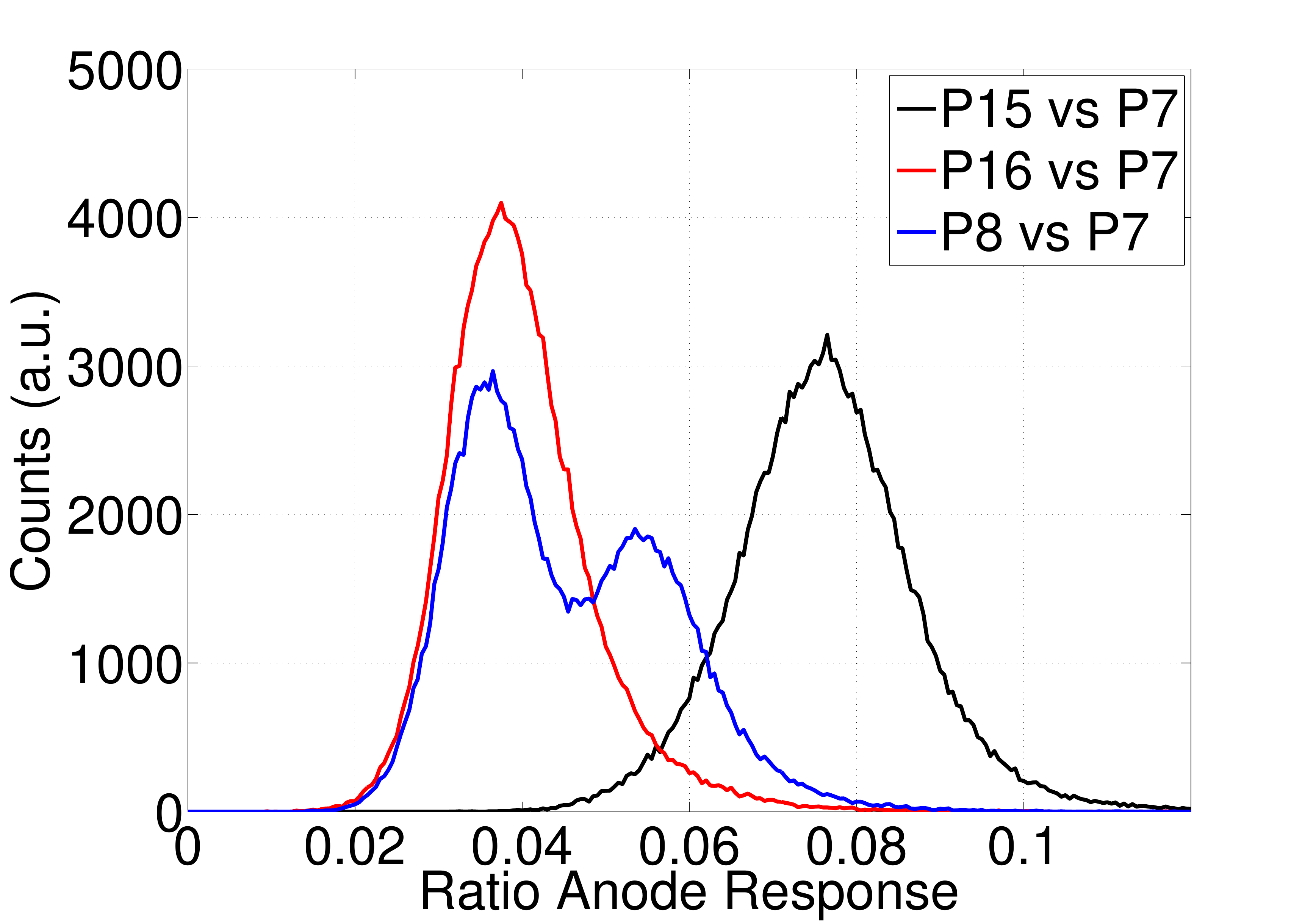}
		\includegraphics[width=0.48\textwidth]{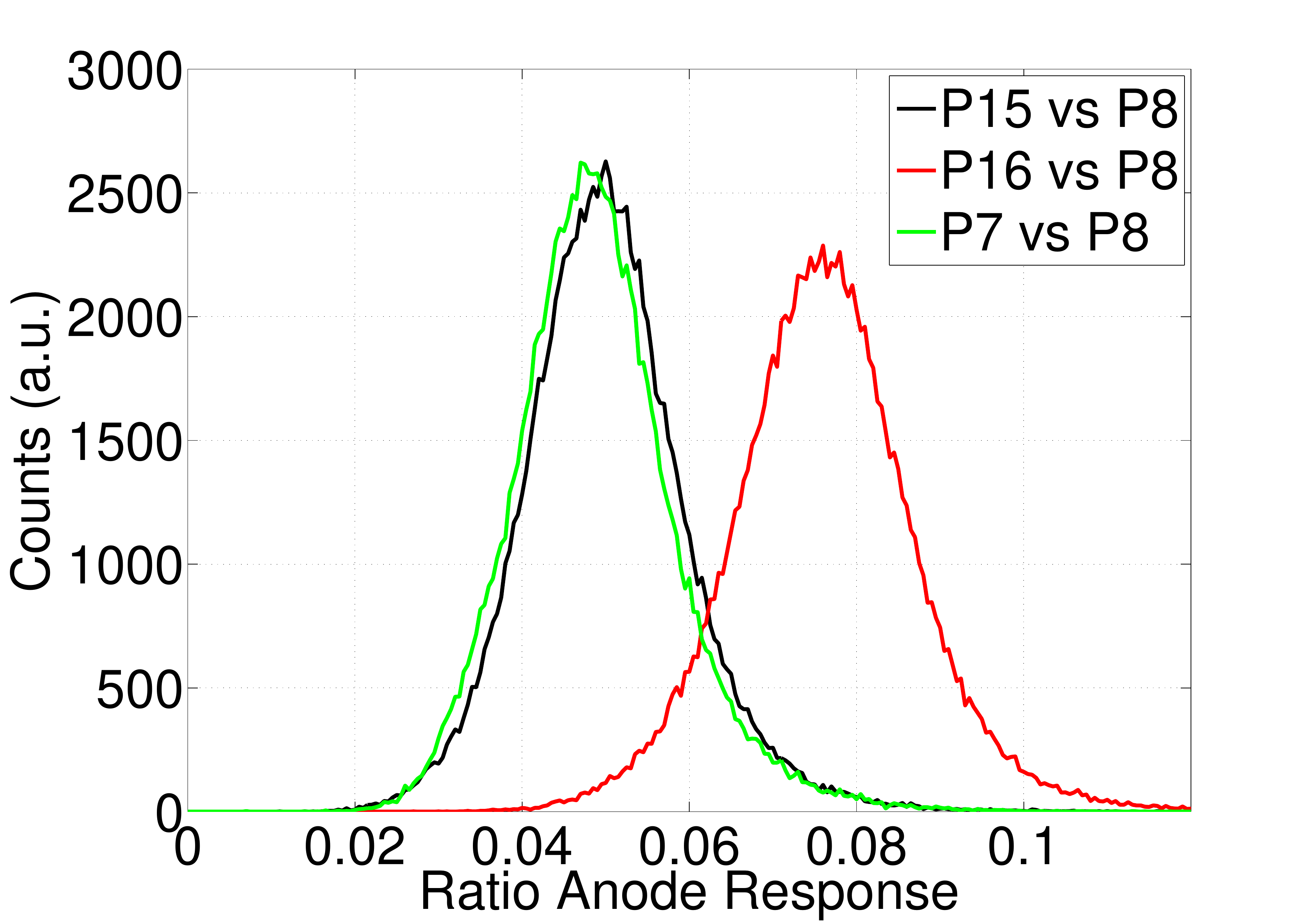}
		\includegraphics[width=0.48\textwidth]{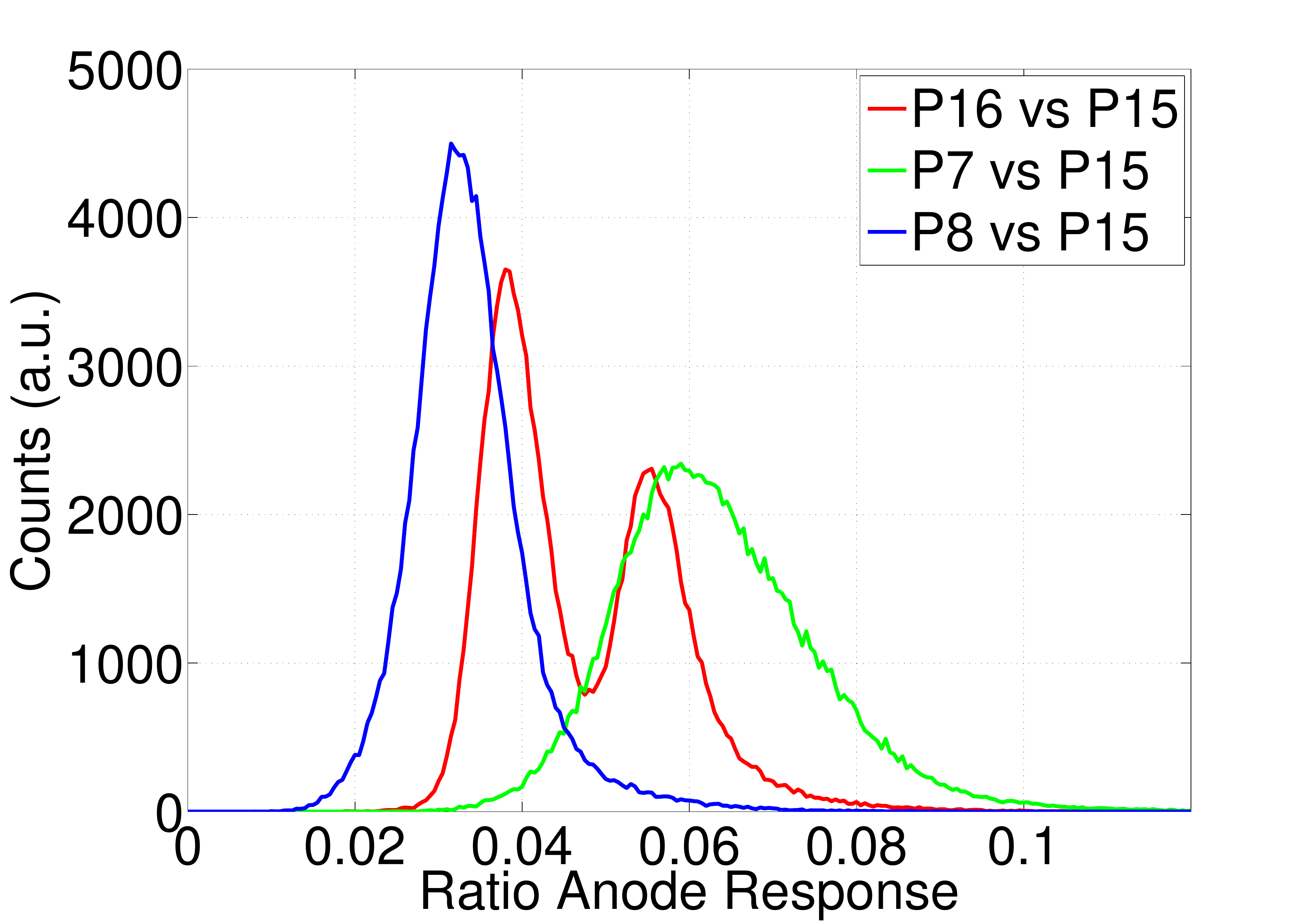}
		\includegraphics[width=0.48\textwidth]{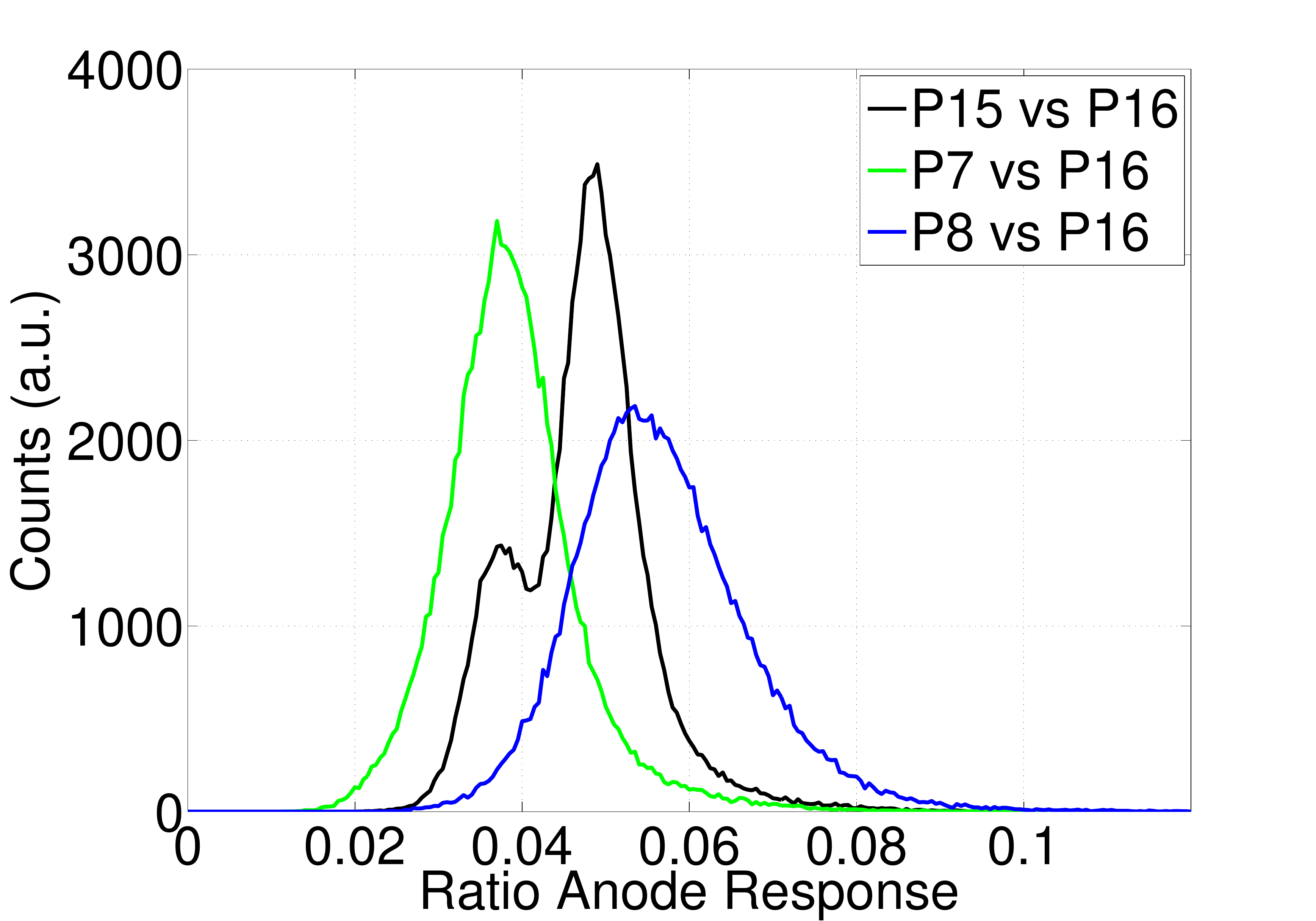}
		\caption{Distribution of the ratio between the amplitude of the cross-talk signals and main signal.}
	\label{fig:cross-talk2}
		\end{minipage}
\end{figure}

A small cross-talk level is required for application in the LHCb RICH, because cross-talk can bias the photon counting and the ring reconstruction. This measurement was performed covering all but one pixels of the MaPMT with a black mask and black tape. Only one pixel was illuminated with the LED and the acquisition was triggered by this channel. In order to avoid the electronic read-out contribution, the input stray capacitance was minimized connecting the MaPMT anodes to the preamplifier with independent, very short coaxial cables. In fig.\ref{fig:cross-talk} a single photon event and the coincident cross-talk events on the neighbouring pixels are plotted. They show a small oscillation with a period of about 4 ns which is probably induced through the dynode bias. The shape of the induced signals and the main signal are very different. Plotting the ratio between the amplitude of the cross-talk signals and the main signal, the cross-talk is estimated to be about 5-10\% , as shown in fig.\ref{fig:cross-talk2}. 


\subsection{Temperature Dependance}
The MaPMT response as a function of temperature has been studied. This is important for correlating the characterization in the laboratory with the final experimental environment. 
The setup to perform these measurements was the following: the MaPMT with its read-out electronics was put in a climatic chamber (Votsch VC 4018), working in the range $-30\,^{\circ}\mathrm{C}$ to $50\,^{\circ}\mathrm{C}$. The behaviour of the electronic read-out chain as a function of the temperature was studied separately, in order to separate its contribution to the measurement. Cherenkov light was simulated with a commercial blue LED located outside of the climatic chamber at constant temperature. Two features of the MaPMT were studied: the single photon spectrum and the rate of dark signals. 

\begin{figure}[h!]
\centering
		\includegraphics[width=0.47\textwidth]{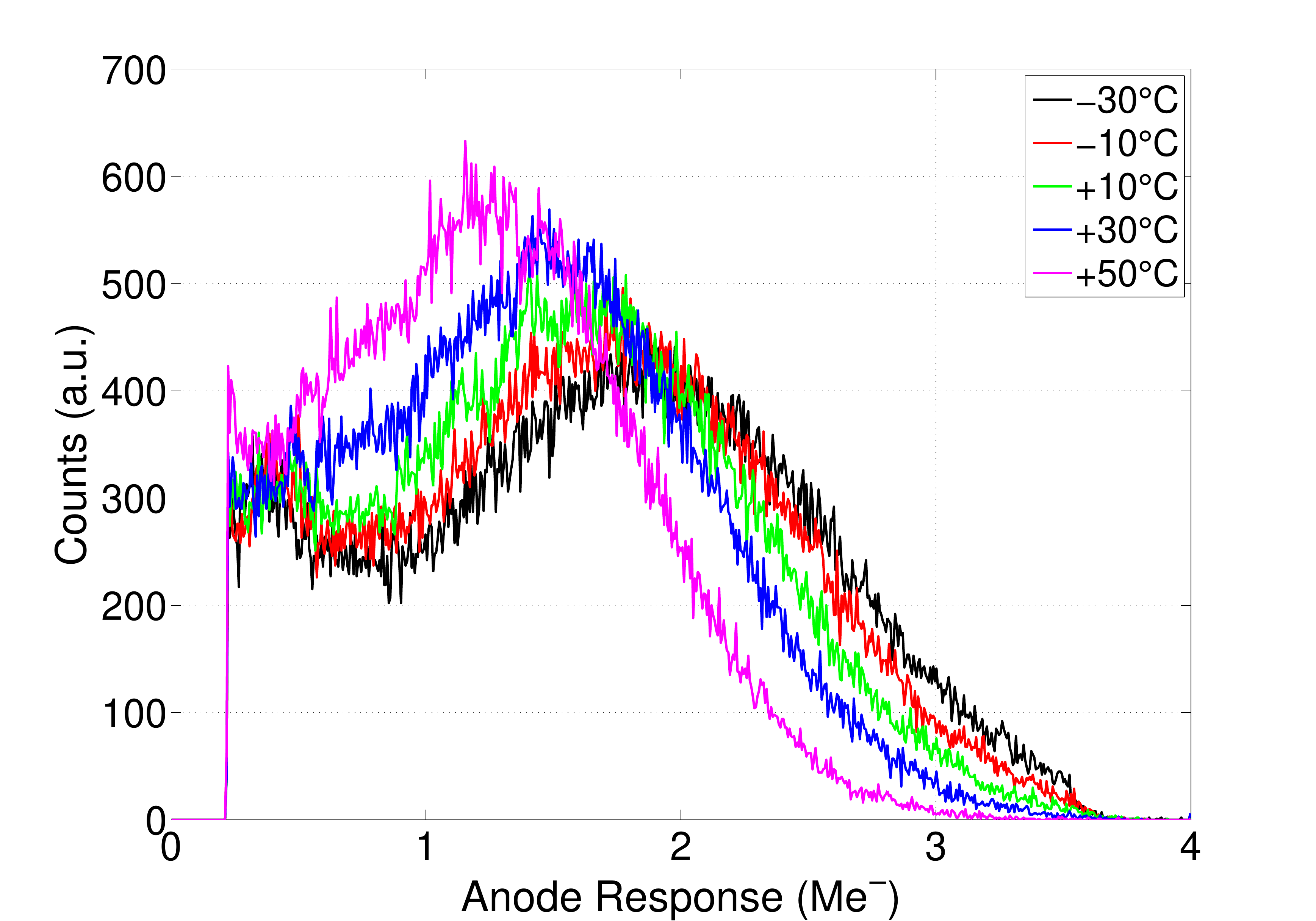}
		\hspace{5mm}%
		\includegraphics[width=0.47\textwidth]{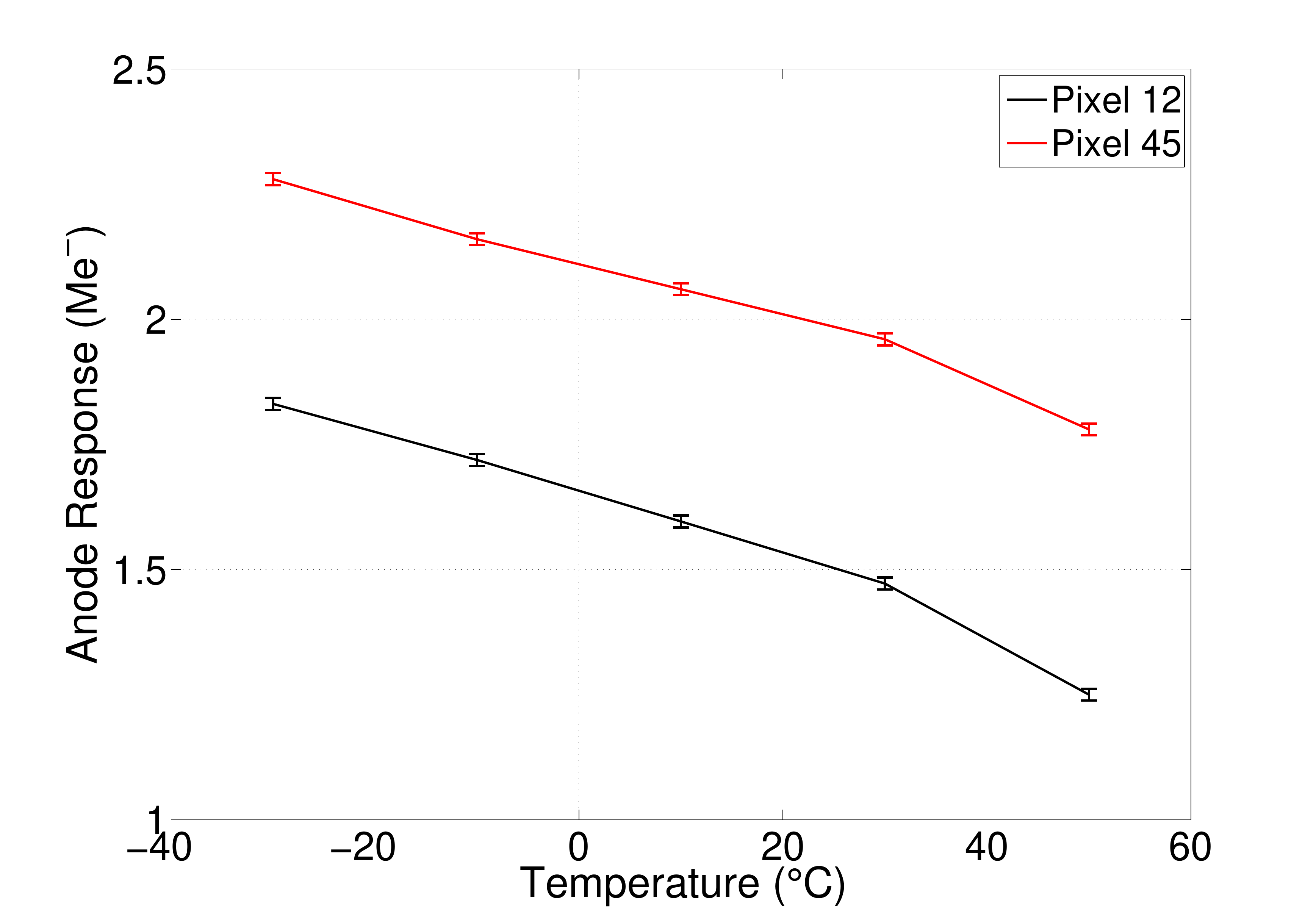}
		\caption{Spectrum of single photon events (left) and gain (right) as a function of temperature (Pixel 12 and Pixel 45, MaPMT SN-ZN0170 biased at -850 V).}
	\label{fig:temperature}
\end{figure}

The measurements show a decrease of the gain increasing the temperature (fig.\ref{fig:temperature}), which can be explained by the reduction of the mean free path of the secondary electrons inside the dynodes. This behaviour is well known and studied (\cite{bibt1}, \cite{bibt2} and \cite{bibt3}) and it can be summarized as follows. The photoelectron energy does not change with temperature, because it is determined only by the electric field between dynodes. Thus, fixing the bias voltage, the electrons hit the dynode with similar energies and predominantly interact with them by electron-electron interactions, which have a negligible dependence on the temperature. Thus, irrespective of the temperature, the incident electrons penetrate the dynode for approximately the same length, exciting the same number of secondary electrons. These particles have a low kinetic energy so that they manly interact with phonons. This interaction has a strong temperature dependence and the mean free path of electrons inside the dynode decreases by increasing the temperature. Therefore, a smaller number of electrons is able to escape from the dynode surface and takes part in following multiplication steps. The global result is a gain decrease. Figure \ref{fig:temperature} shows that the gain depends on the temperature quite linearly and the reduction slope is of about 5500 $e^-/^{\circ}\mathrm{C}$. Assuming $30\,^{\circ}\mathrm{C}$ as reference value, a gain variation of about 0.3\% per $^{\circ}\mathrm{C}$ is expected.

\begin{figure}[h!]
\centering
		\includegraphics[width=0.47\textwidth]{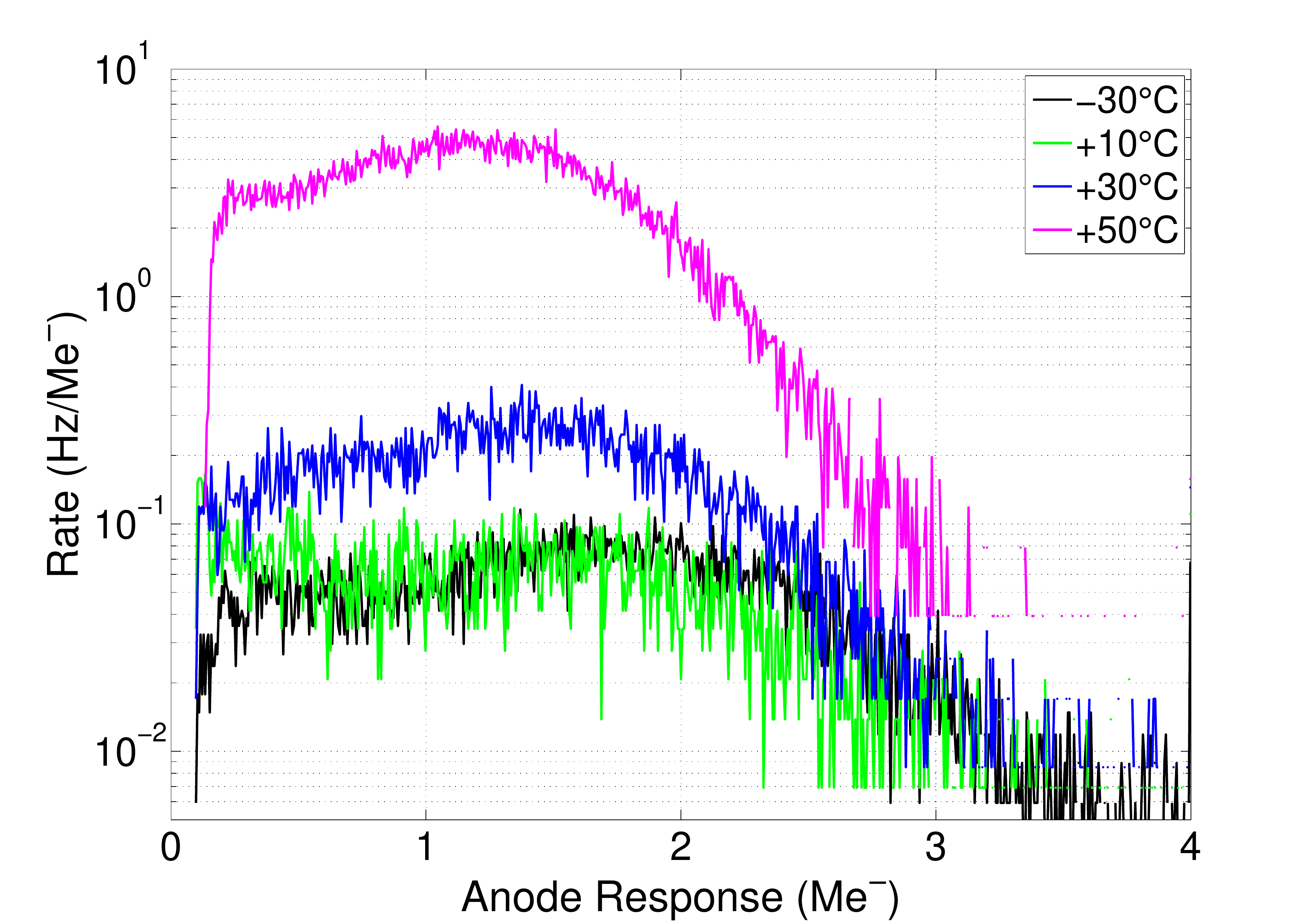}
		\hspace{5mm}%
		\includegraphics[width=0.47\textwidth]{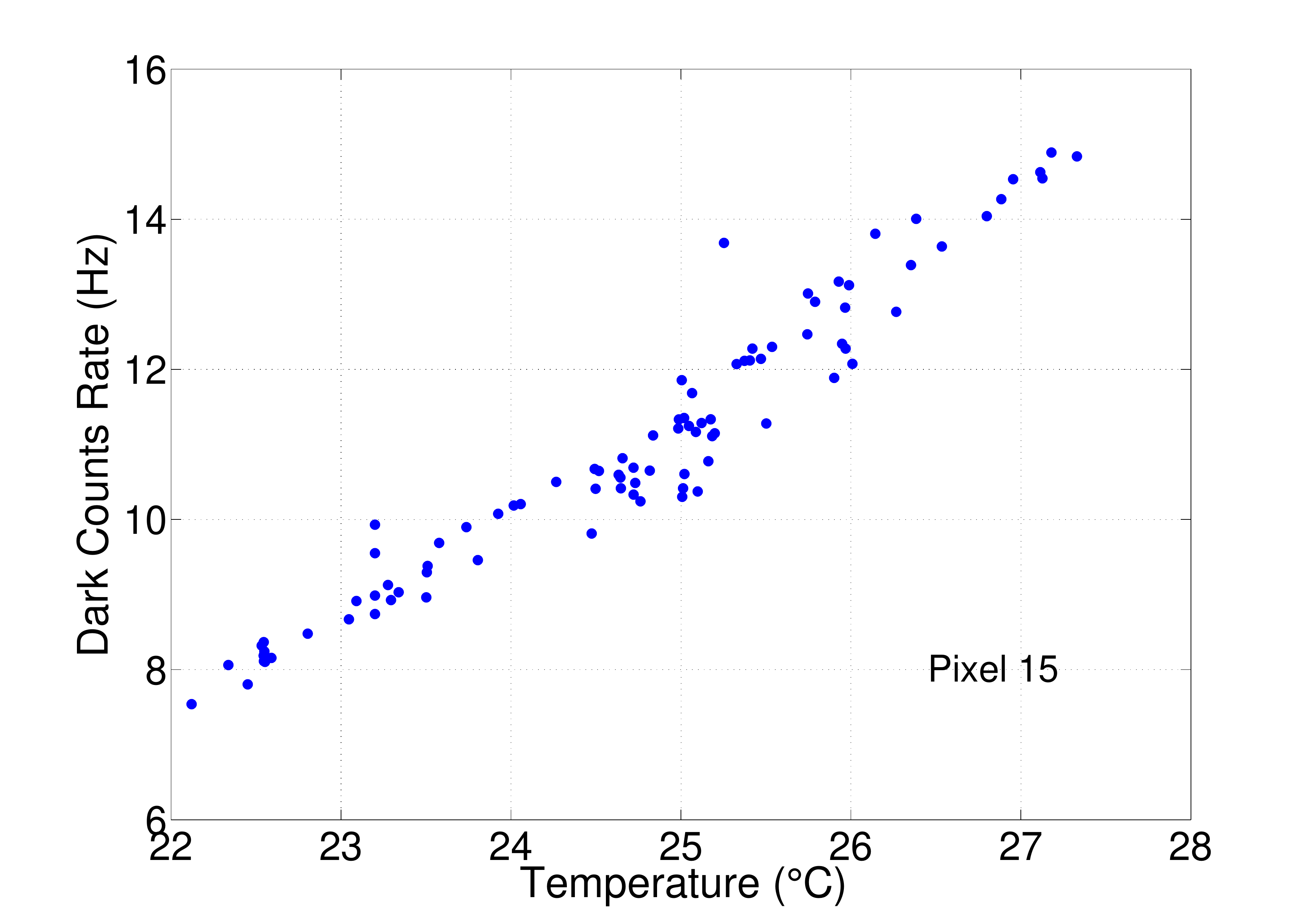}
		\caption{Spectrum (left) and rate (right) of dark signals as a function of temperature (Pixel 12 and Pixel 15, MaPMT SN-ZN0170 biased at -850 V).}
	\label{fig:temperaturenoise}
\end{figure}

The second feature studied is the temperature dependence of the dark current (fig.\ref{fig:temperaturenoise}). The rate of noise signals increases greatly with temperature (the dark signal rate at $50\,^{\circ}\mathrm{C}$ is about 30 times larger than the one at  $-30\,^{\circ}\mathrm{C}$). The reason of this phenomenon is that, increasing the temperature, the number of electrons which have enough thermal energy to escape from the dynodes or the photocathode surface and generate a multiplication process increases. In first approximation, at room temperature the dark event rate increases quite linearly with the temperature, as it can be seen in fig.\ref{fig:temperaturenoise} (right). 

\section{Behaviour in critical environment condition} \label{sec:Environment}

The performance of the \itshape R11265-103-M64 \upshape tube was investigated also in case it has to operate in critical environment conditions, such as under the action of a magnetic field or during long time operation at intense light exposure. These tests are particularly important in case the \itshape R11265-103-M64 \upshape will be employed in application such as the LHCb RICH detector where a fringe magnetic field up to 30~G is expected and the photosensor is supposed to withstand the intense illumination of the most central zone for about ten years. In the next paragraphs the investigations about the behaviour of the \itshape R11265 \upshape as a function of a longitudinal magnetic field up to 100~G are shown. In the last section a quantitative estimation of the aging effects (gain loss, photocathode efficiency deterioration and dark current increase) is illustrated.

\subsection{Behaviour in magnetic field}
If the photomultipliers have to work in magnetic field it is important to know their behaviour and eventually be shielded with a high permeability material (such as Skudotech$\textregistered$ or  MuMetal$\textregistered$).

\subsubsection{Setup}
\begin{figure}[h!]
	\centering
			\includegraphics[width=0.45\textwidth]{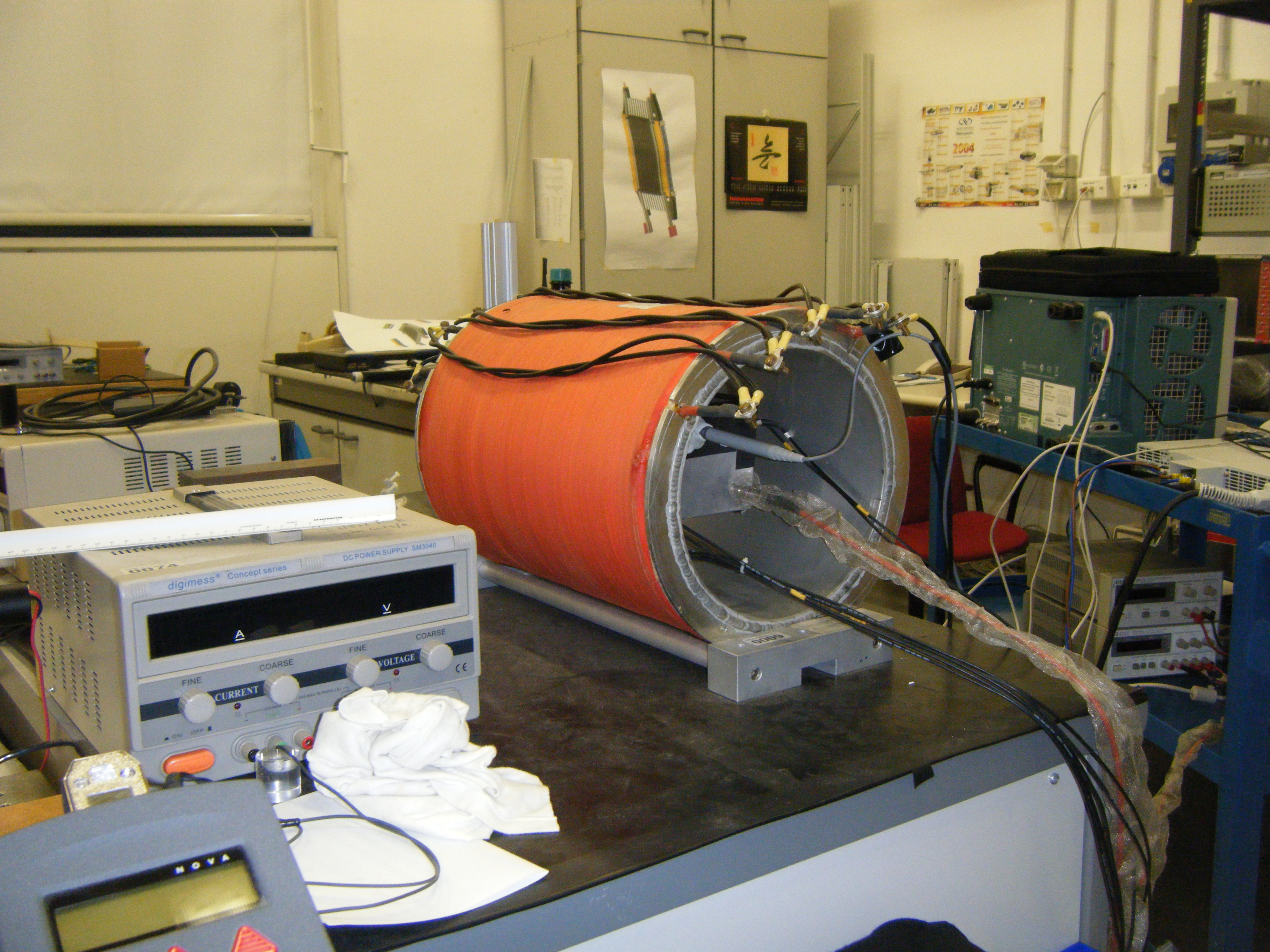}
			\caption{A picture of the setup used for the tests of the \itshape R11265 \upshape MaPMT in magnetic field.}
			\label{fig:magneticsetup}
\end{figure}

The response of photomultiplier can be sensitive to a magnetic field because it might induce some electrons to change their trajectory from one dynode to the following one.
As far as a transversal magnetic field (perpendicular to the tube axis) is concerned, the metallic layer covering the MaPMT lateral surface behaves  by itself as a shield, thus the effects are usually negligible up to 100~G and more.
The main effects are induced by a longitudinal magnetic field (parallel to the tube axis), as already demonstrated in  \cite{bib6}.
The studies were made with a \itshape R11265-103-M64 \upshape MaPMT placed in a longitudinal magnetic field produced by a solenoid (fig.\ref{fig:magneticsetup}).
An automatic system, made of custom MATLAB$\textregistered$ scripts, changed the current flowing in the solenoid in order to obtain fields ranging from 25~G to 100~G.
The field magnitude was measured with a Hirst GM04 gaussmeter.

A commercial blue LED, which was biased a very low voltage so that it operated in single photon region, illuminated the 32 tested pixels. 
The MaPMT was biased at -1050 V at an average gain of about 2.5~$Me^-/$photon. 
The MaPMT signal was amplified using a classic charge sensitive preamplifier circuit and finally acquired by a DT5720 CAEN Desktop Digitizer\footnote{The DT5720 is a 4 channel 12 bit 250 MS/s Desktop Waveform Digitizer with 2 Vpp single ended input. The DC offset is adjustable via a 16 bit DAC on each channel in the $\pm$ 1V range.} fixing a threshold of  $\sim$60~ke$^-$.
Although all the measurements were performed at room temperature, with a maximum range of $1^{\circ}$C, small fluctuations are expected between different acquisitions. For this reason in order to check the system stability, three zero field observations were performed at the beginning, in the middle, and at the end of the run. 

\subsubsection{Measurements without the magnetic shield}\label{NoMagneticShield}
\begin{figure}[h!]
	\centering
		\includegraphics[width=0.45\textwidth]{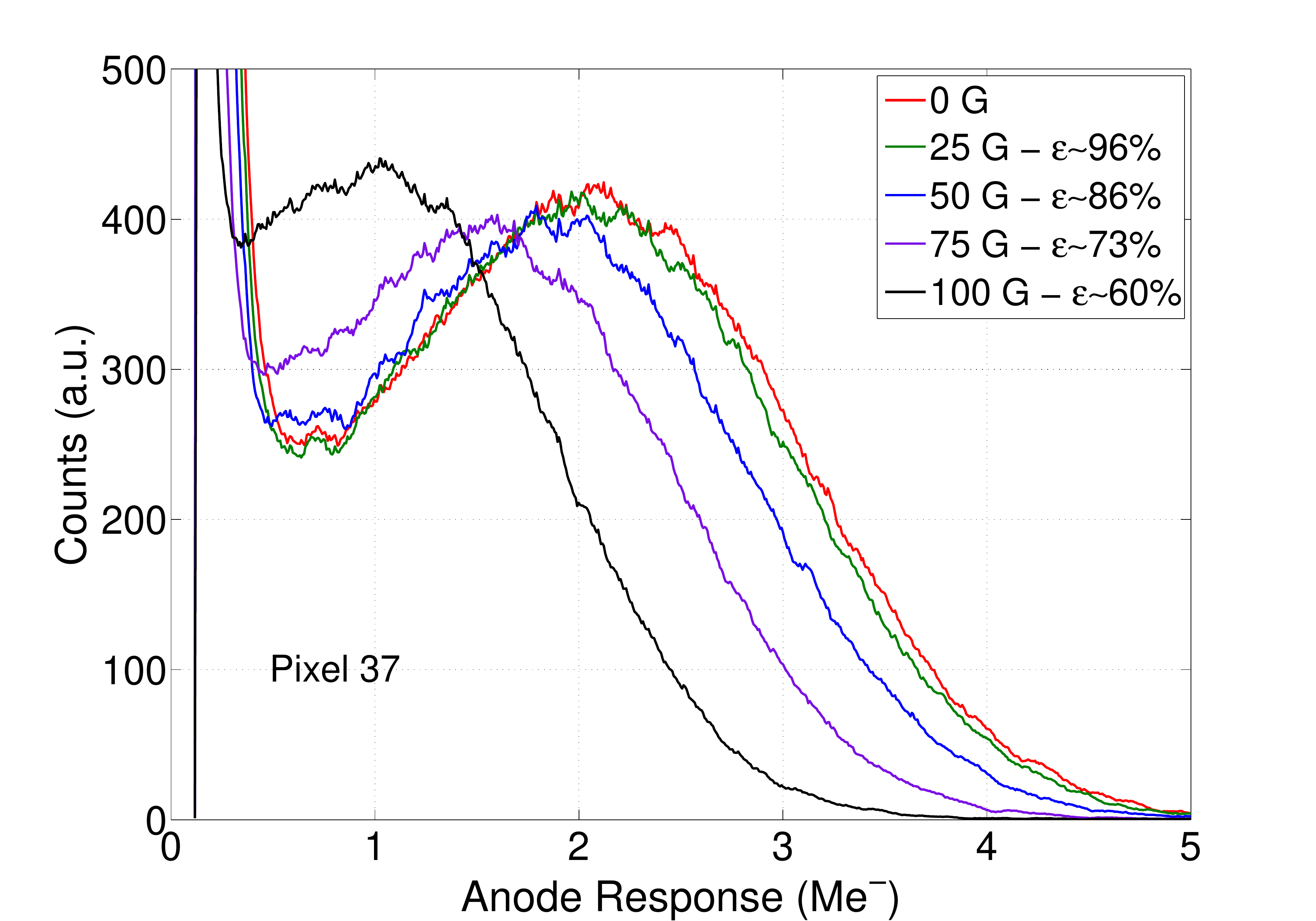}
		\hspace{5mm}%
		\includegraphics[width=0.45\textwidth]{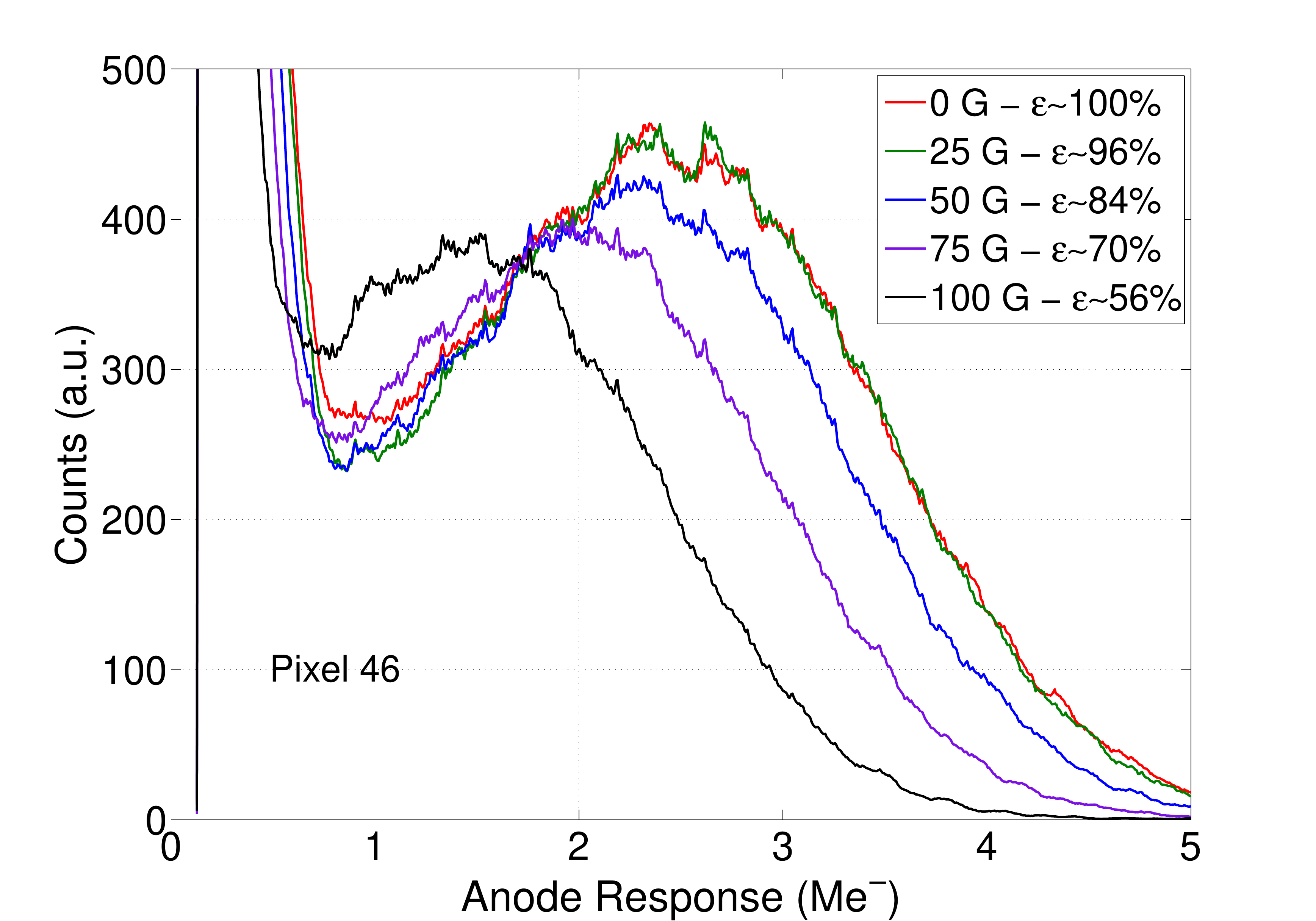}
		\includegraphics[width=0.45\textwidth]{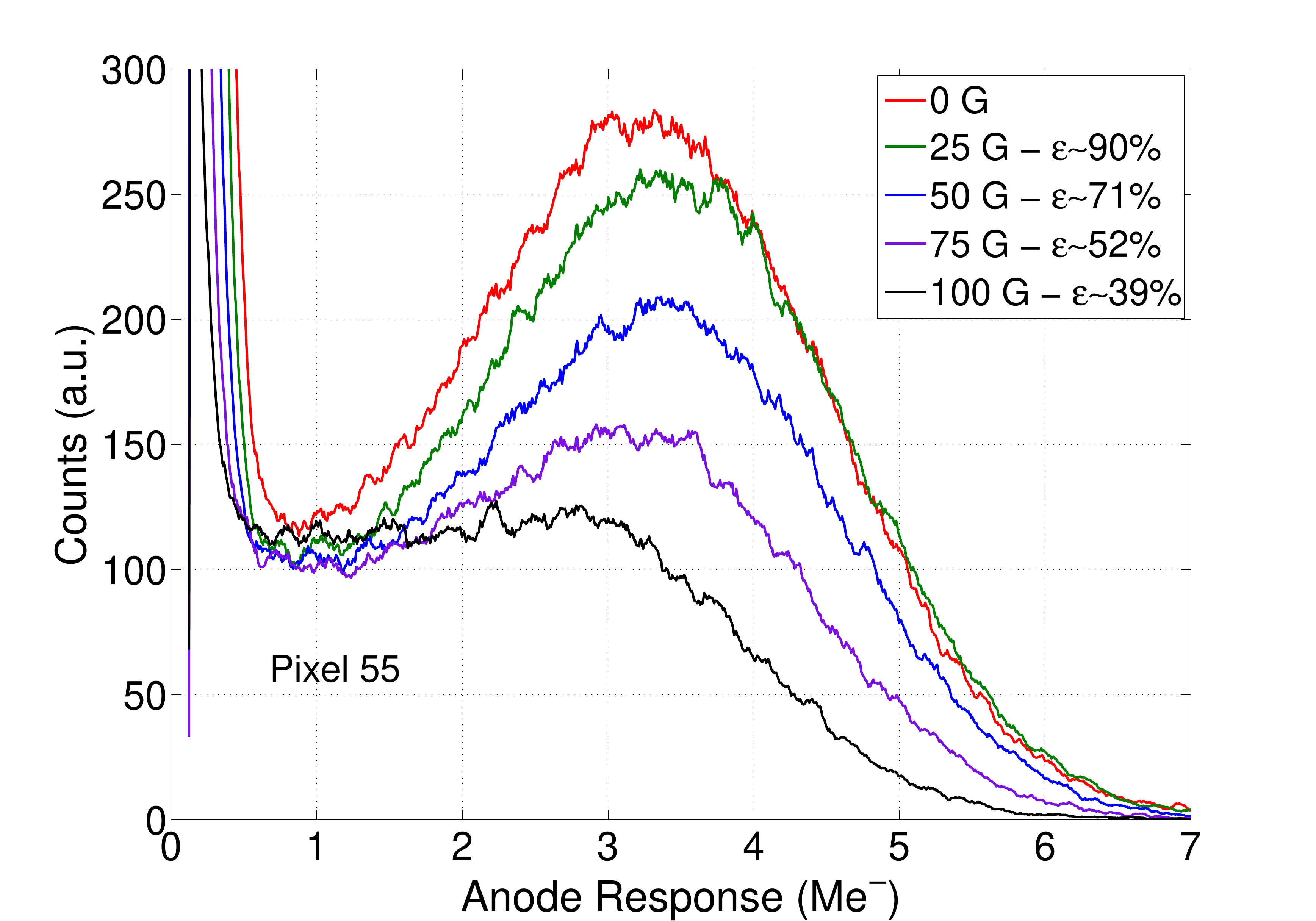}
		\hspace{5mm}%
		\includegraphics[width=0.45\textwidth]{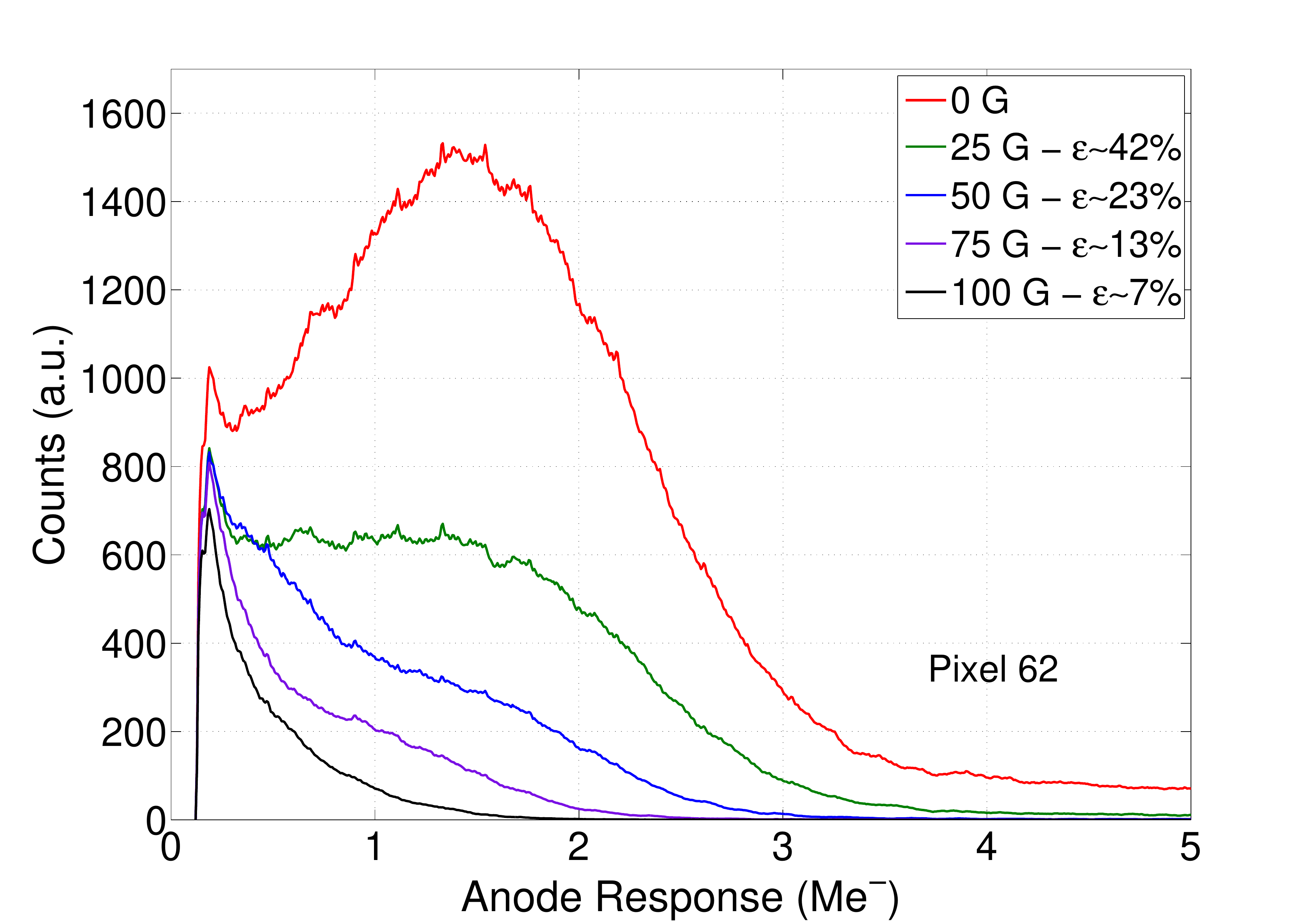}
		\caption{Each graph shows the superposition of single photon spectra acquired under the action of different longitudinal magnetic fields values. The two upper plots refer to pixels located in the central part of the MaPMT while the two lower plots correspond to pixels in the peripheral region (MaPMT SN-ZN0702 biased at -1050V.)}
	\label{fig:MagneticNoShield}
\end{figure}

The spectra acquired for 4 pixels are shown in fig.\ref{fig:MagneticNoShield}. The central pixels turn out to be quite insensitive to a weak magnetic field, while significant single photon spectra deformation is observed for the peripheral pixels. 
\begin{figure}[h!]
	\centering
		\includegraphics[width=0.45\textwidth]{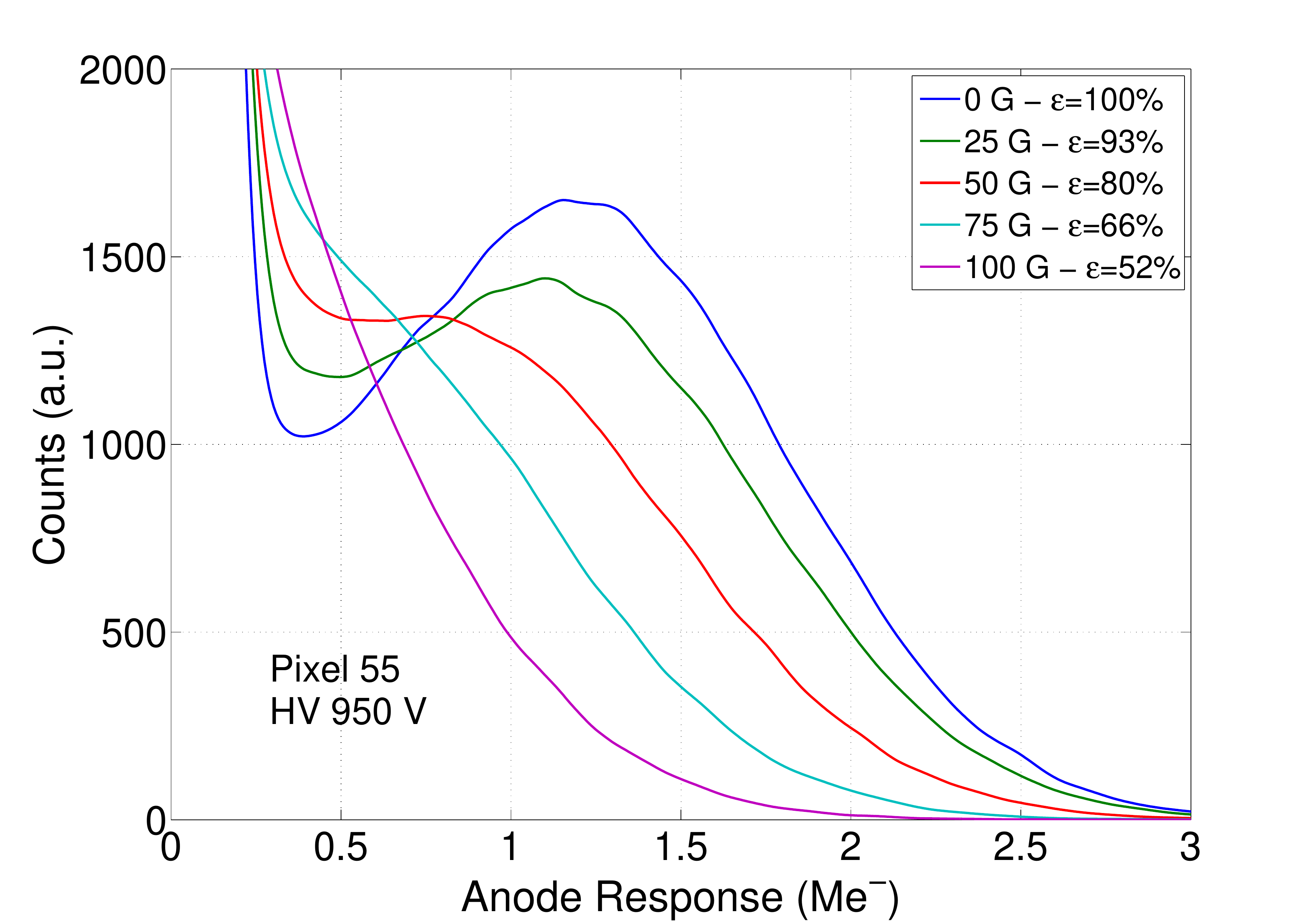}
		\hspace{5mm}%
		\includegraphics[width=0.45\textwidth]{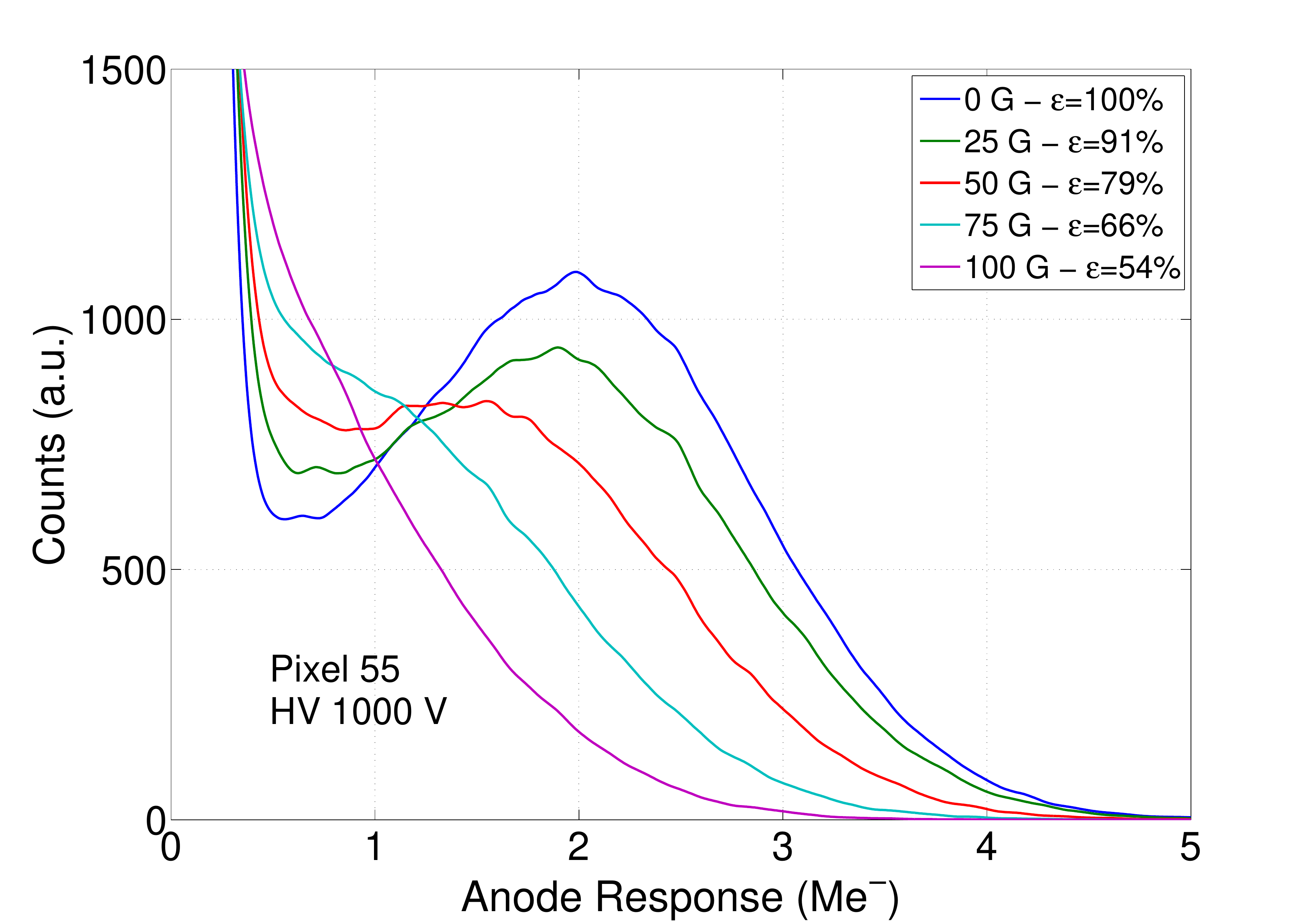}
		\includegraphics[width=0.45\textwidth]{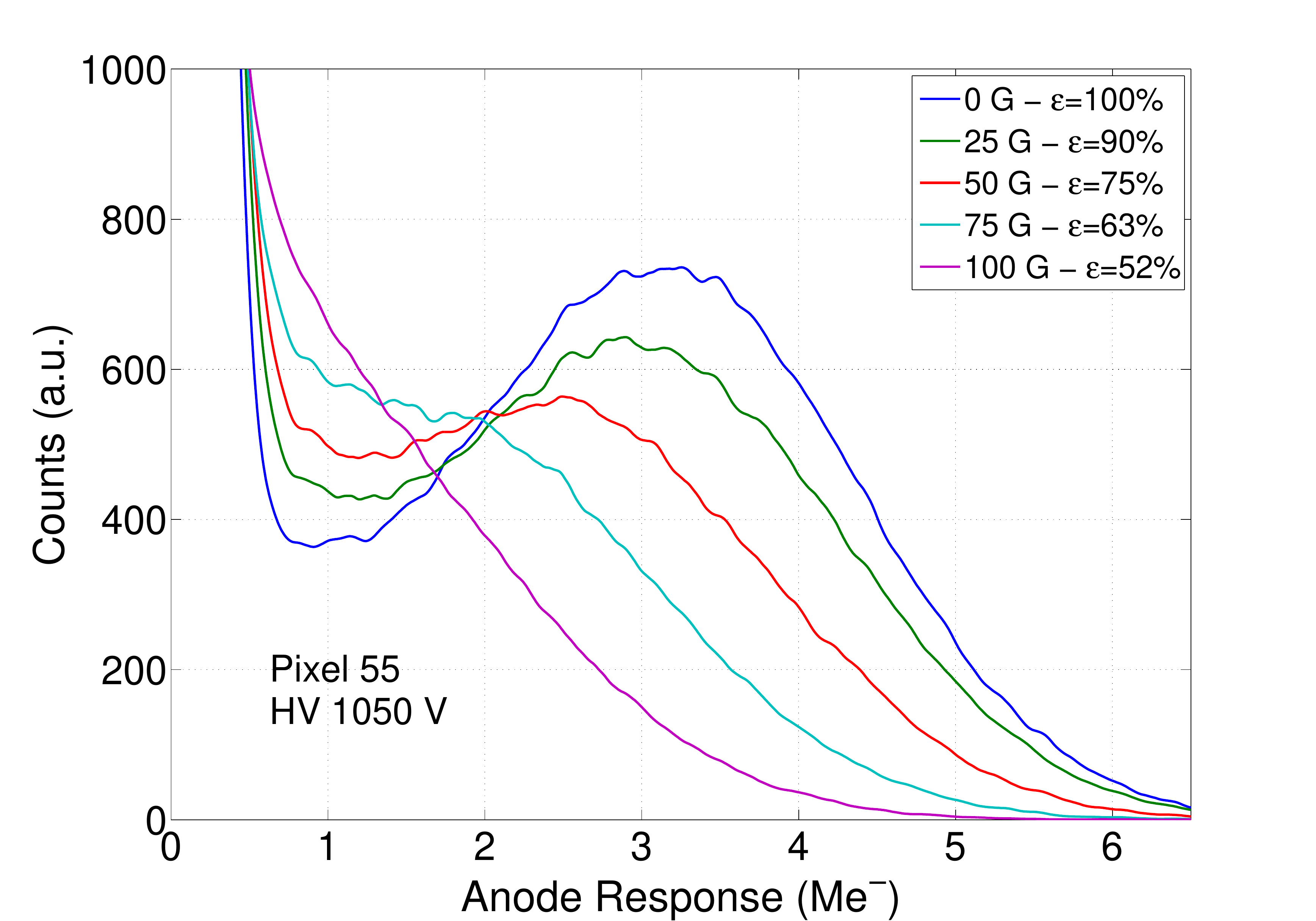}
		\hspace{5mm}%
		\includegraphics[width=0.45\textwidth]{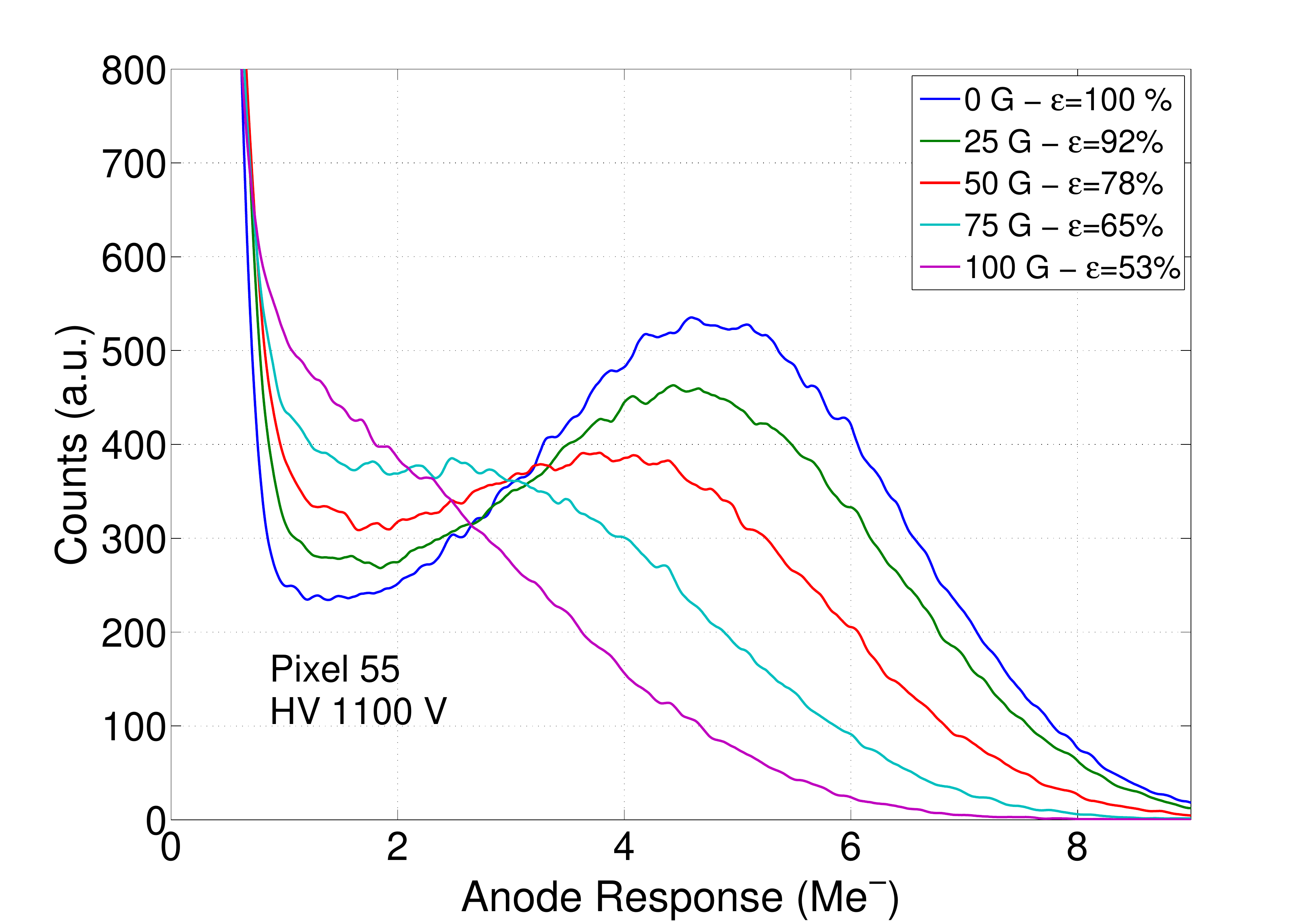}
		\caption{Each graph shows the superposition of single photon spectra acquired under the action of different longitudinal magnetic fields values. All the plots refer to the pixel 55 located near the side of the device. The data were acquired biasing the MaPMT (FA0019) at a bias voltage that ranges from -950 V to -1100 V moving from the top-left picture to the bottom-right one.}
	\label{fig:FieldVsVoltage}
\end{figure}

If we define the efficiency $\epsilon$ as the rate of events whose amplitude is larger than the noise level (fixed at $0.2$~$Me^-$), then for the most external pixel (pixel 62 in fig.\ref{fig:MagneticNoShield}) the efficiency at 25 Gauss is less than half with respect to the one measured without magnetic field. Considering a more intense field, large distortion can also be observed for central pixels, while for the peripheral ones the single photon peak almost merges with the pedestal. The results suggest that these devices need to be shielded to work properly in a magnetic field. 

Note that this effect was found to be almost independent on the bias voltage 
and on the average gain of the MaPMT. As it can be observed in fig.\ref{fig:FieldVsVoltage}, no significant variation of the photon detection efficiency is visible increasing the bias voltage from -950 V to -1100 V. 

\subsubsection{Tests and optimization of shielding configuration}
\begin{figure}[h!]
	\centering
		\includegraphics[width=0.45\textwidth]{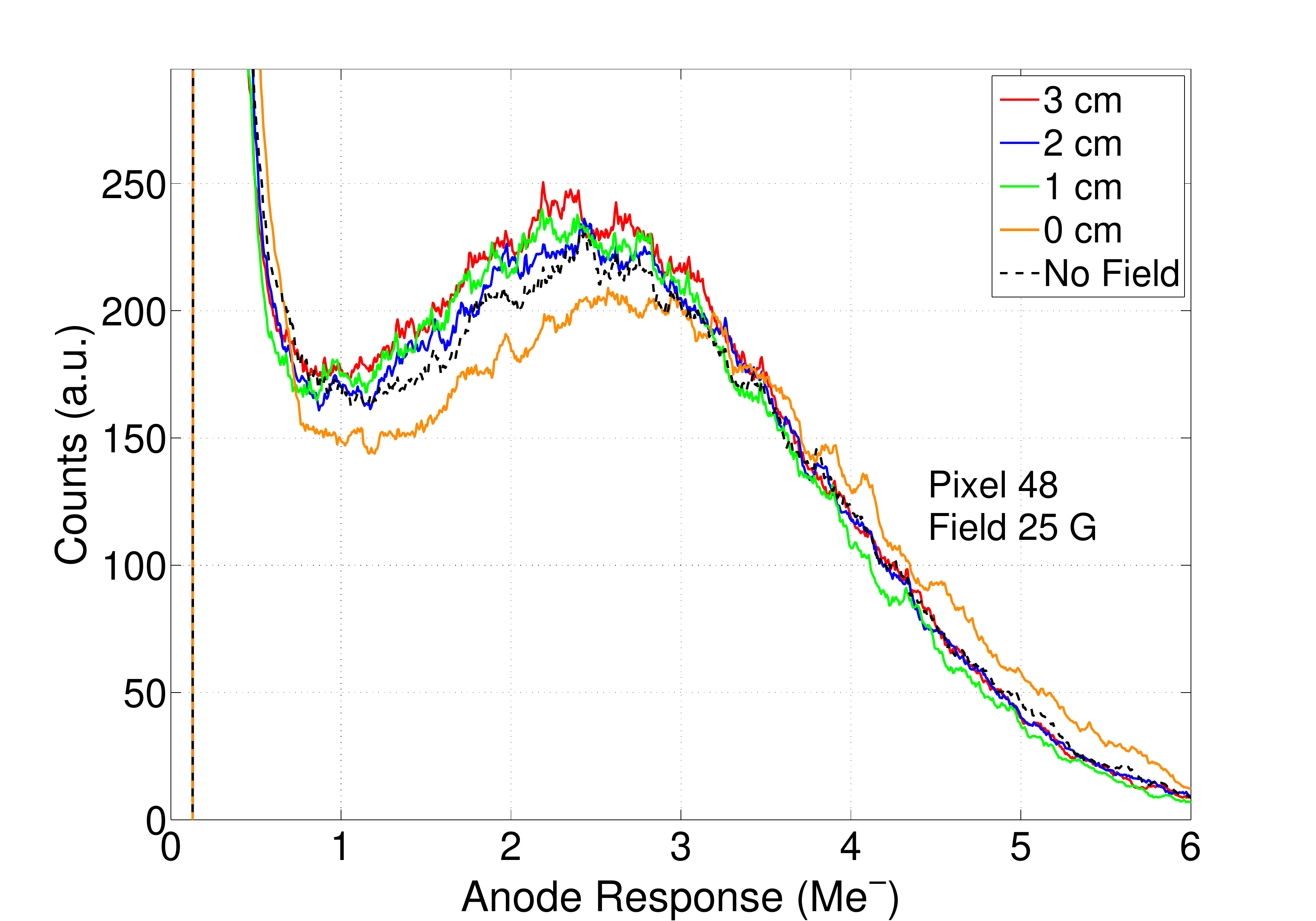}
		\hspace{5mm}%
		\includegraphics[width=0.45\textwidth]{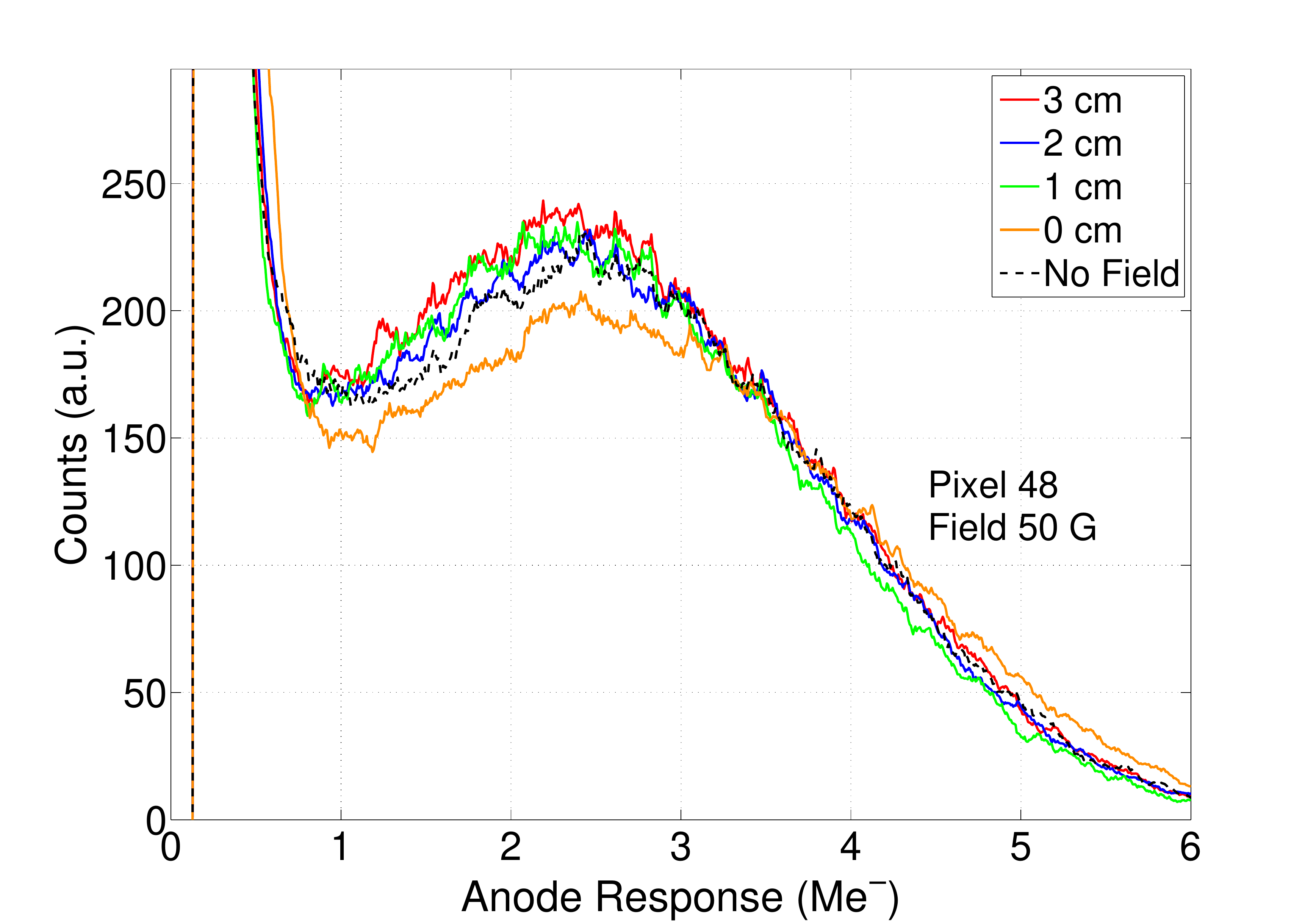}
		\includegraphics[width=0.45\textwidth]{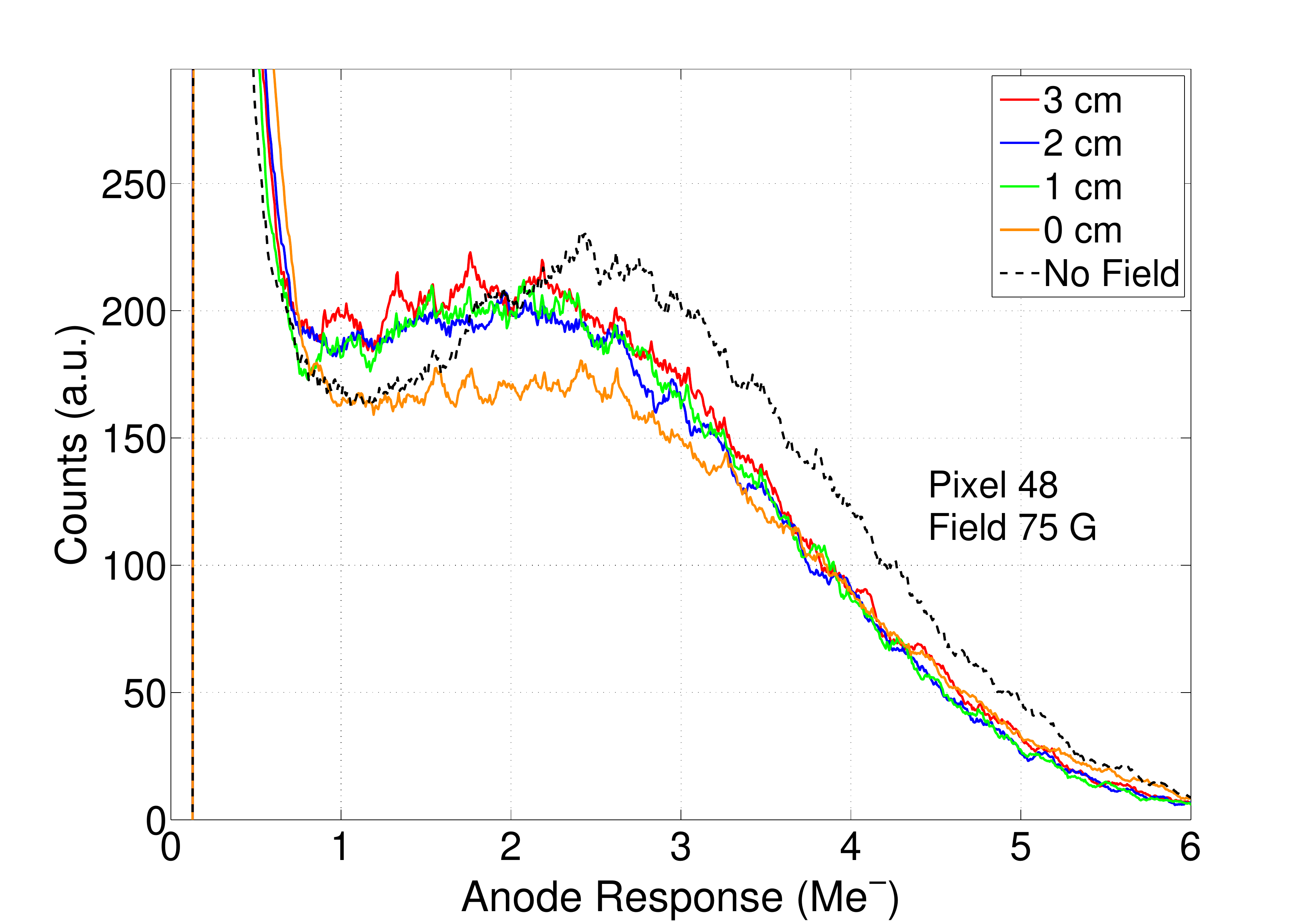}
		\hspace{5mm}%
		\includegraphics[width=0.45\textwidth]{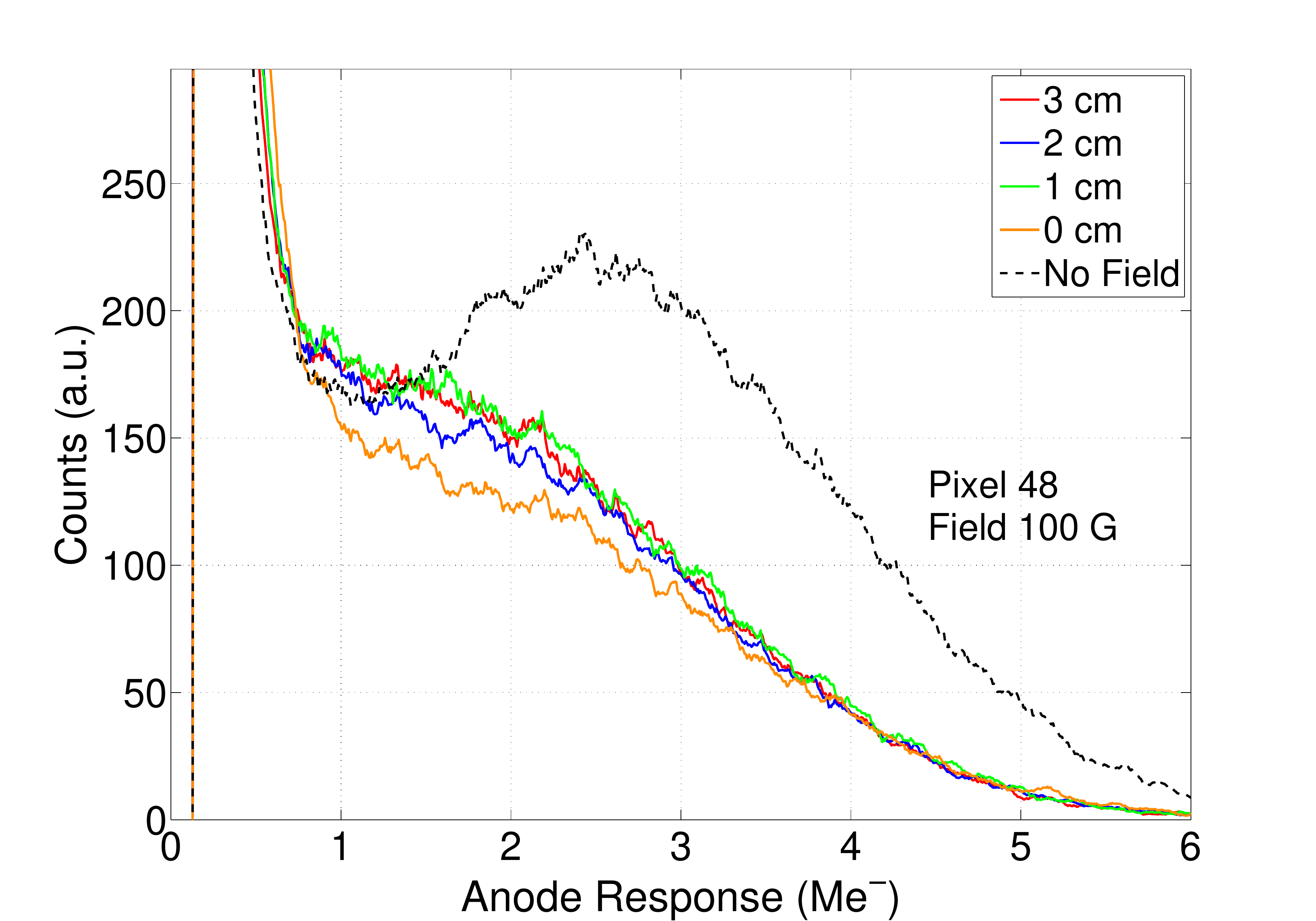}	
		\caption{Each graph shows the superposition of single photon spectra acquired with the MaPMT \itshape R11265-103-M64 \upshape (ZN0702) under the action of a longitudinal magnetic field using shields which differ in the protrusion length. All the plots refer to the pixel 48, located on the side of the device. The magnetic field ranges from 25~G to 100~G and increases moving from the top-left to the bottom-right graph.}
	\label{fig:ProtrudePixel48}
\end{figure}

In light of the results shown in the previous paragraph \ref{NoMagneticShield}, a magnetic shielding is necessary to preserve a good response at the peripheral pixels. 
The main parameters which have to be considered to obtain an effective shielding are its material, the geometrical shape, the thickness and the length of its protrusion from the photocathode surface. 
In order to absorb a static magnetic field, high permeability materials are needed. 

For all the measurements described in this section, the shields were made of 200 $\mu$m thick Skudotech$\textregistered$ (produced by SELITE), material with a nominal maximum permeability\footnote{Further informations: http://www.bmtel.it/Skudotech.pdf} of $3.27\cdot10^5$. 
This material was chosen also for its flexibility, so different shapes can be easily obtained and the thickness of the shield can be increased by wrapping multiple Skudotech$\textregistered$ layers around the tube.
Other high permeability materials, like the MuMetal$\textregistered$ (maximum magnetic permeability value\footnote{Further informations: http://www.mushield.com/material-specs.shtml} $\sim 2\cdot10^5$), can be appropriate shields. 
By the way, the MuMetal$\textregistered$ is used around the HPD detectors currently employed in the LHCb RICH. 
In all the measurements, the  Skudotech$\textregistered$ sheet was wrapped around the lateral surface of the MaPMT and protruded, by different lengths, from the photocathode window. 

\begin{figure}[h!]
	\centering
		\includegraphics[width=0.45\textwidth]{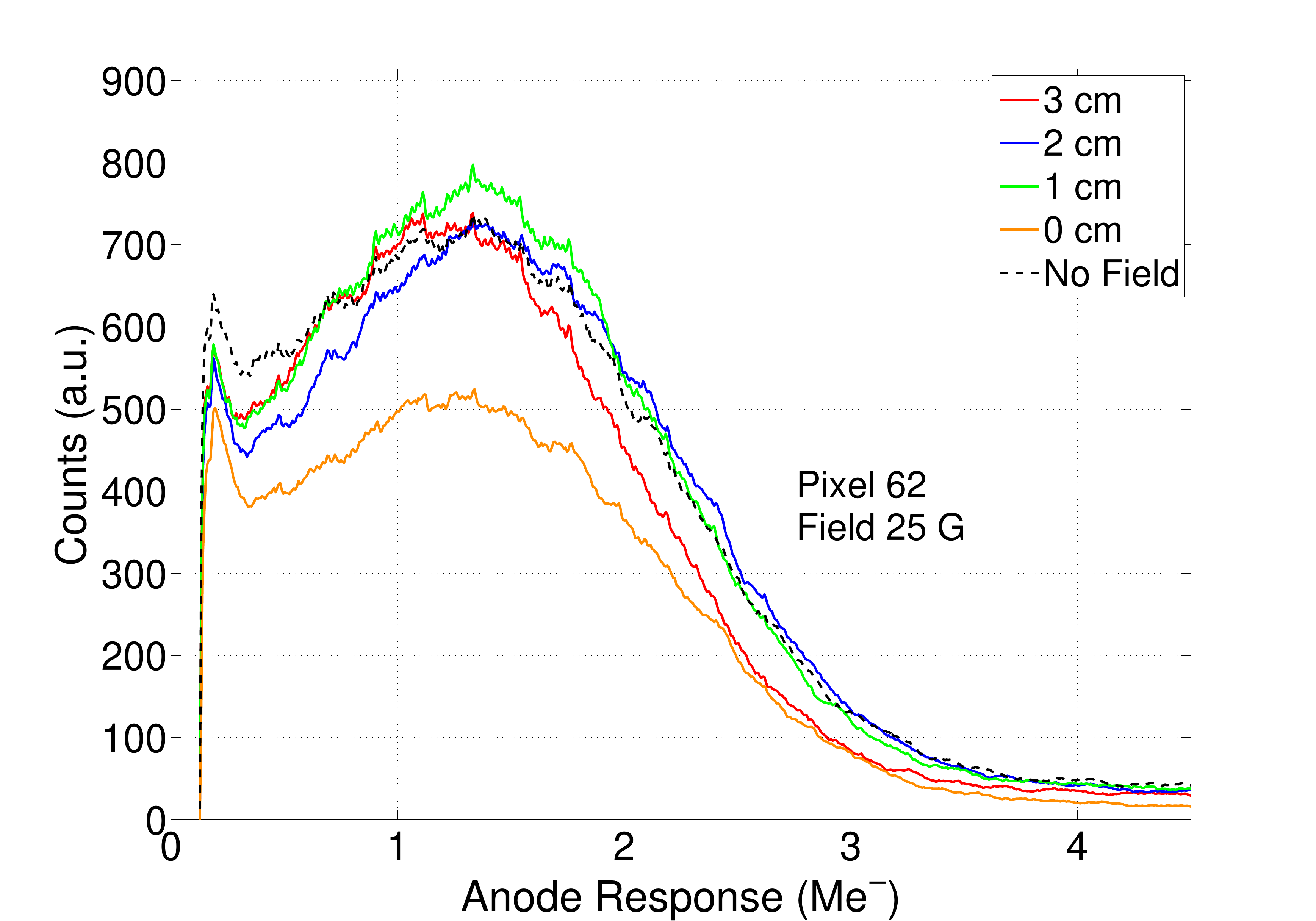}
		\hspace{5mm}%
		\includegraphics[width=0.45\textwidth]{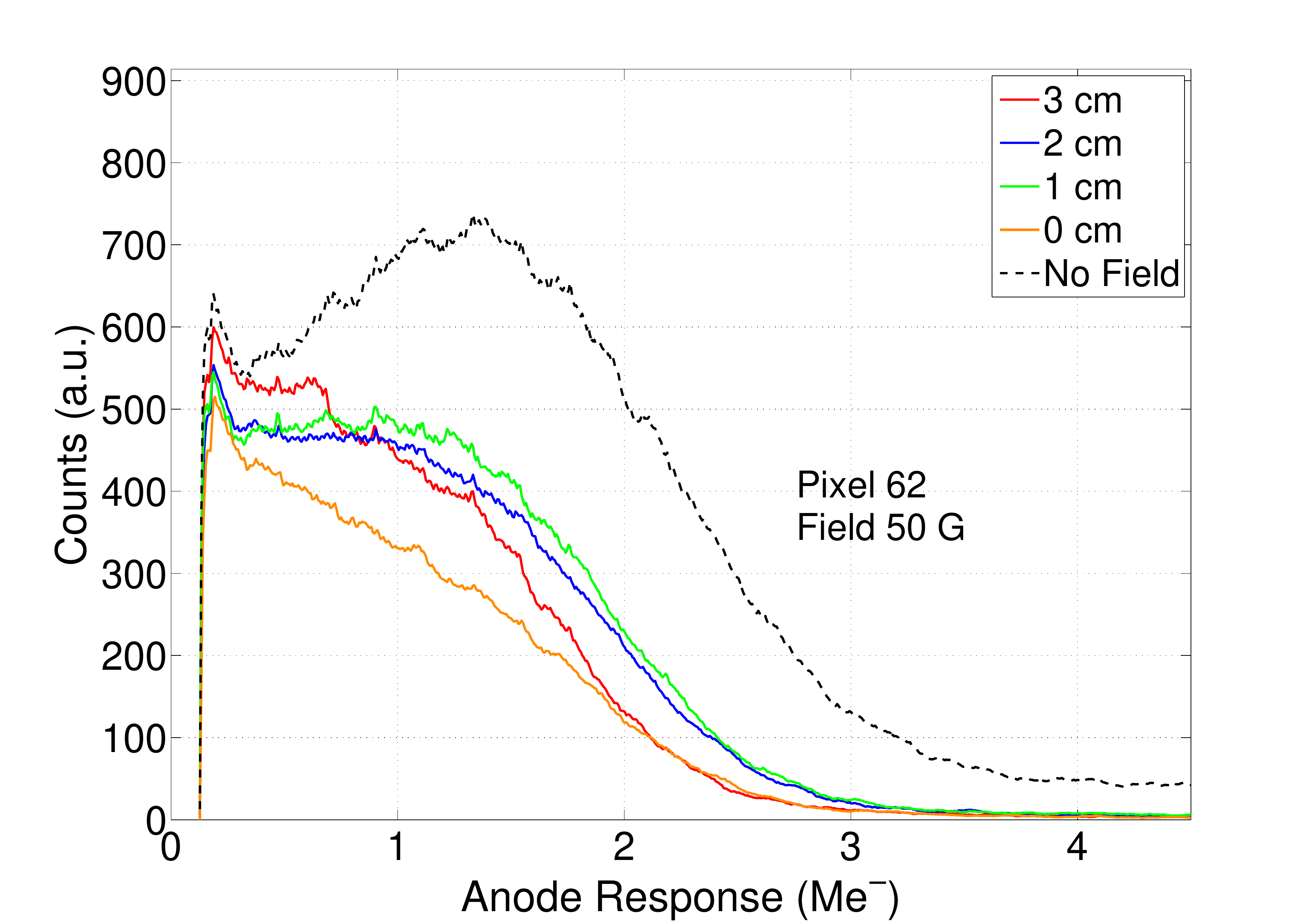}
		\includegraphics[width=0.45\textwidth]{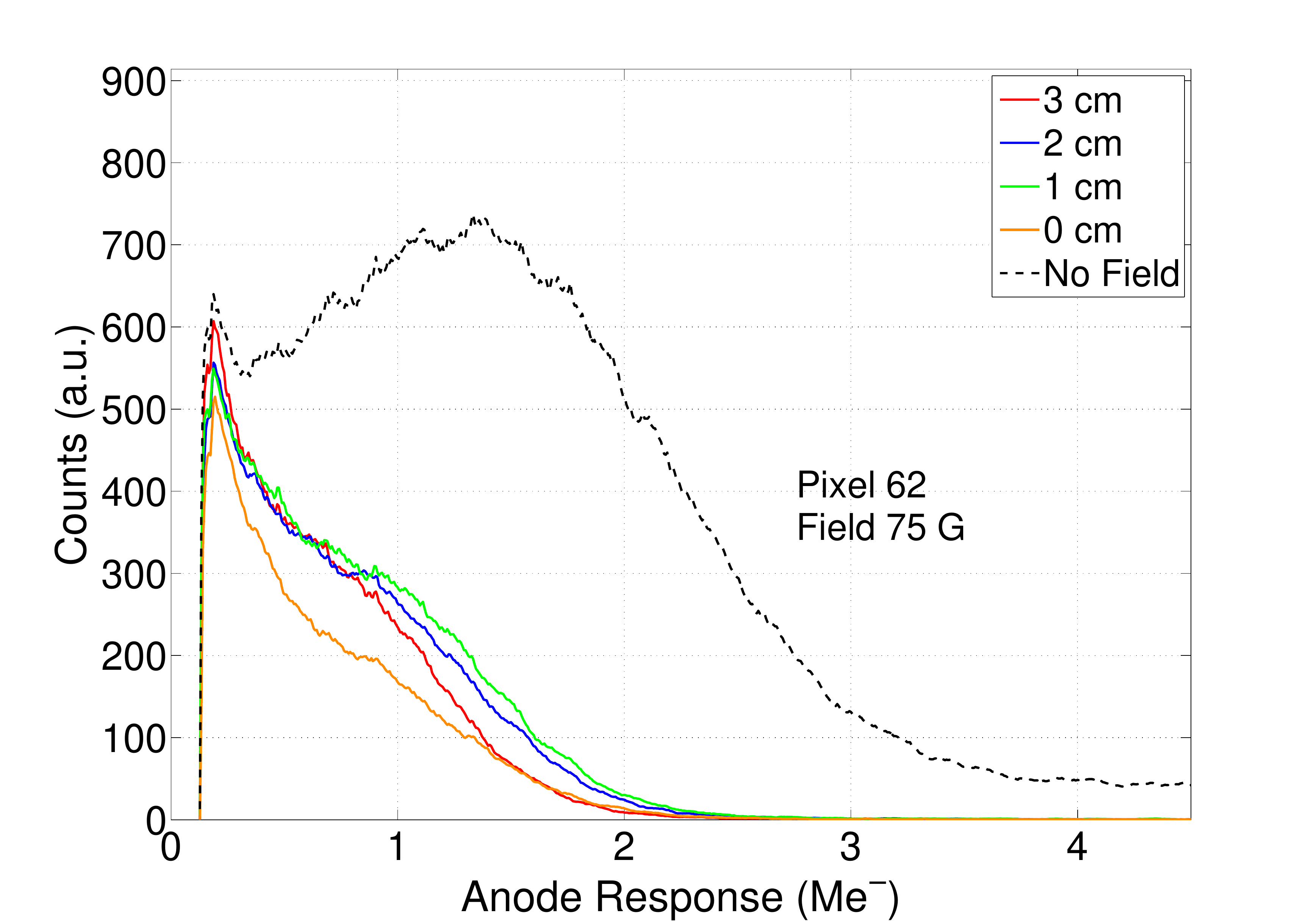}
		\hspace{5mm}%
		\includegraphics[width=0.45\textwidth]{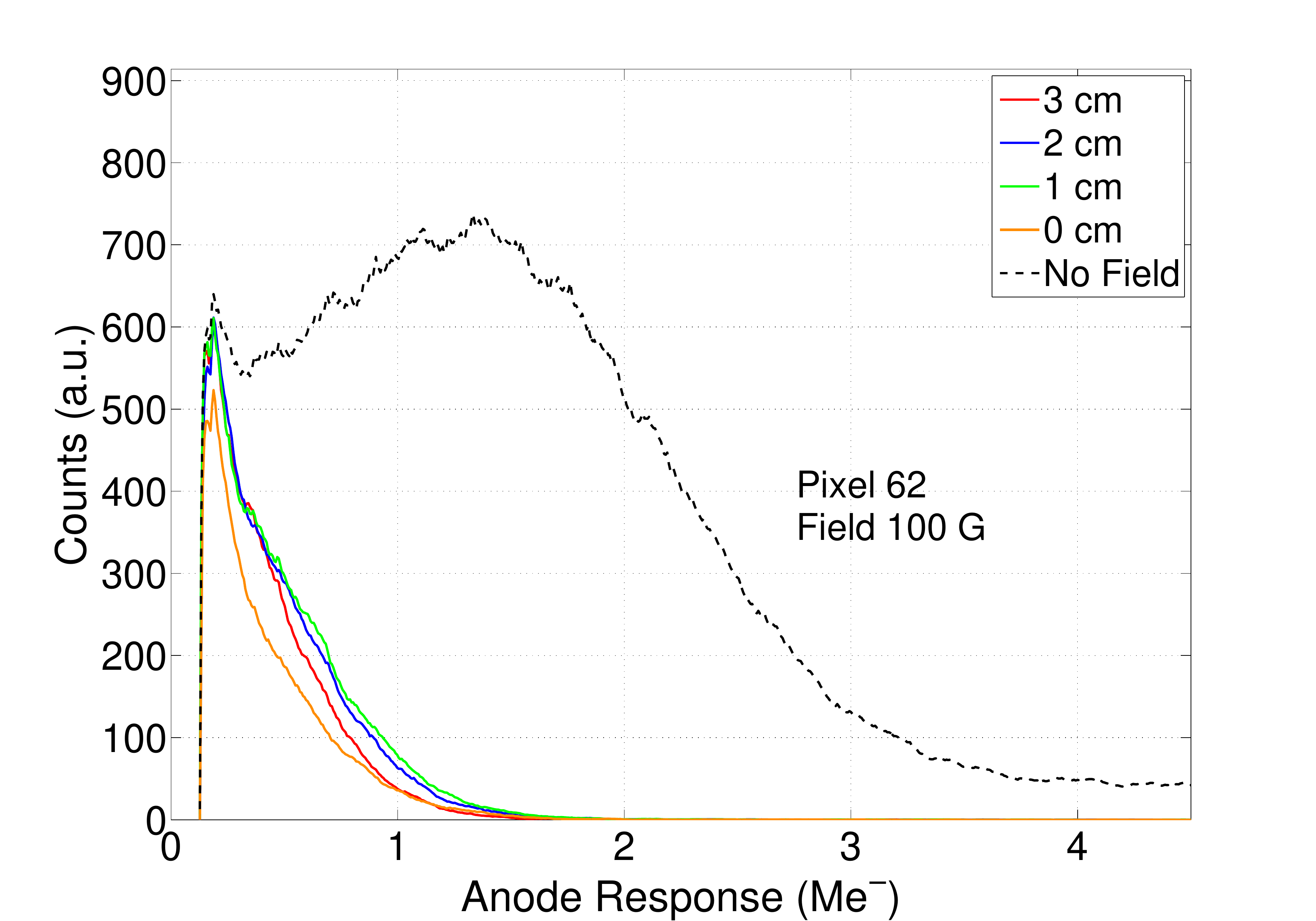}	
		\caption{Each graph shows the superposition of single photon spectra acquired with the MaPMT \itshape R11265-103-M64 \upshape (ZN0702) under the action of a longitudinal magnetic field using shields which differ in the protrusion length. All the plots refer to the pixel 62, located on the side of the device and next to the bias voltage pins. The magnetic field ranges from 25~G to 100~G and increases moving from the top-left to the bottom-right graph.}
	\label{fig:ProtrudePixel62}
\end{figure}

The wrapping did not cover the tube pins at the back.
The thickness and the protrusion lengths are the parameters to be optimized. 
In principle the more the shield protrudes, the more it is effective. 
The drawback is that the protrusion limits the angular acceptance of the photons to be detected.  
Figures \ref{fig:ProtrudePixel48} and \ref{fig:ProtrudePixel62} show the effects of shields with different protrusion lengths. Note that all these shields were made by a single wrap of 200~$\mu$m thick Skudotech$\textregistered$, so they differ only in their length. 

Figure \ref{fig:ProtrudePixel48} shows the single photon spectra acquired on the pixel 48 (located on the side of the MaPMT) as a function of the longitudinal magnetic field, from 25~G to 100~G.  
The shield which does not protrude from the surface (0 cm) is not able to properly absorb the magnetic field and its effectiveness is inadequate even at 25~G. 
All the other shields are more effective and show similar performance. 
In particular, the spectra distortions are critical only in case of magnetic field larger than 50~G. 
Figure \ref{fig:ProtrudePixel62} shows the same results for the pixel 62, located just next to the bias voltage pins. 
In this case, the magnetic field can even turn the pixel efficiency to zero, since the single photon peak is not resolved from the pedestal above 50~G. 
Although both these pixels, 48 and 62, are located on the side of the device, their performance is quite different. 
In order to recover the signal in each pixel, properly shielding all of them, the worst cases must be taken into account. 
For the central pixels, the effects of the magnetic field are much less dramatic.

\begin{figure}[h!]
	\centering
		\includegraphics[width=0.45\textwidth]{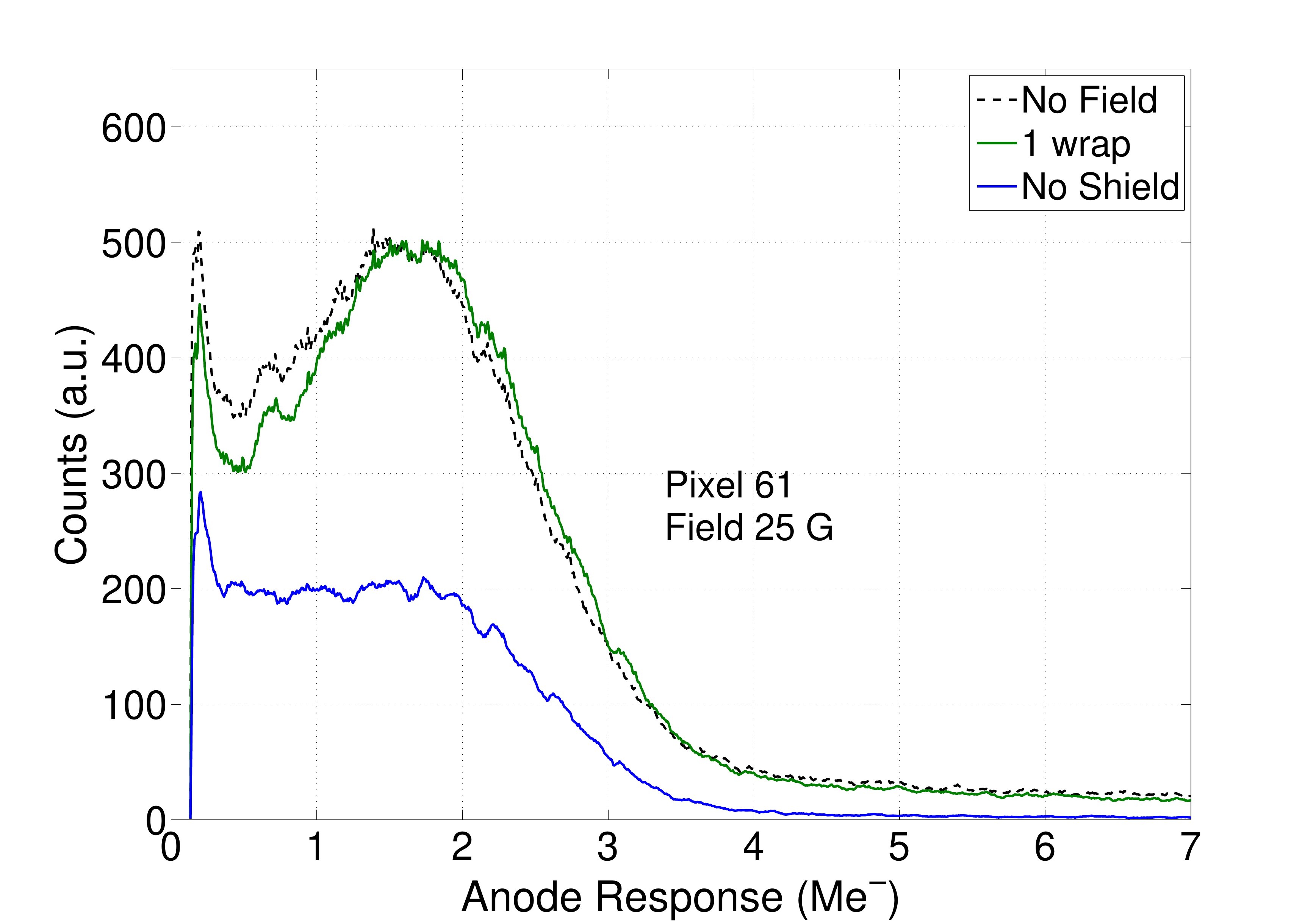}
		\hspace{5mm}%
		\includegraphics[width=0.45\textwidth]{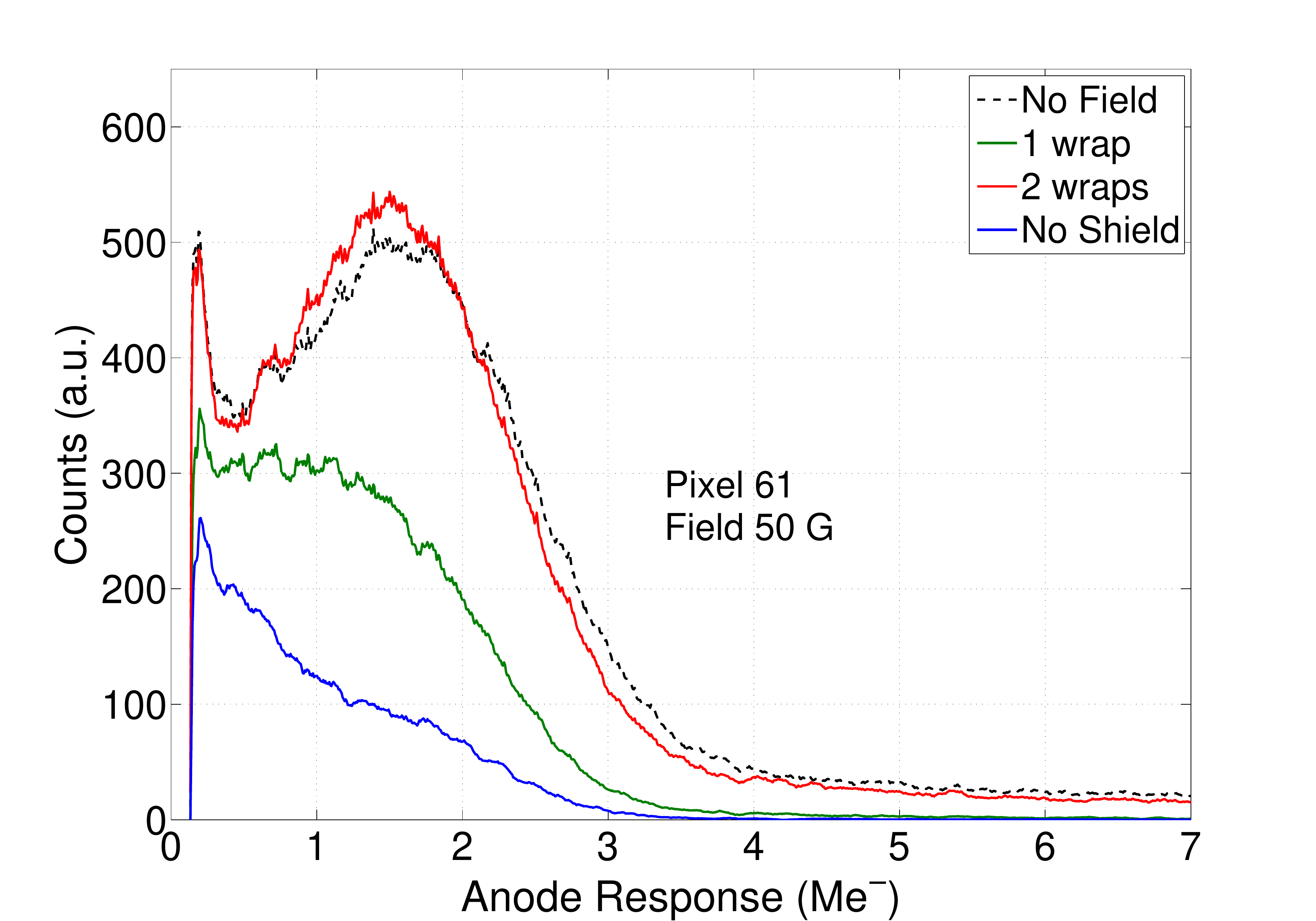}
		\includegraphics[width=0.45\textwidth]{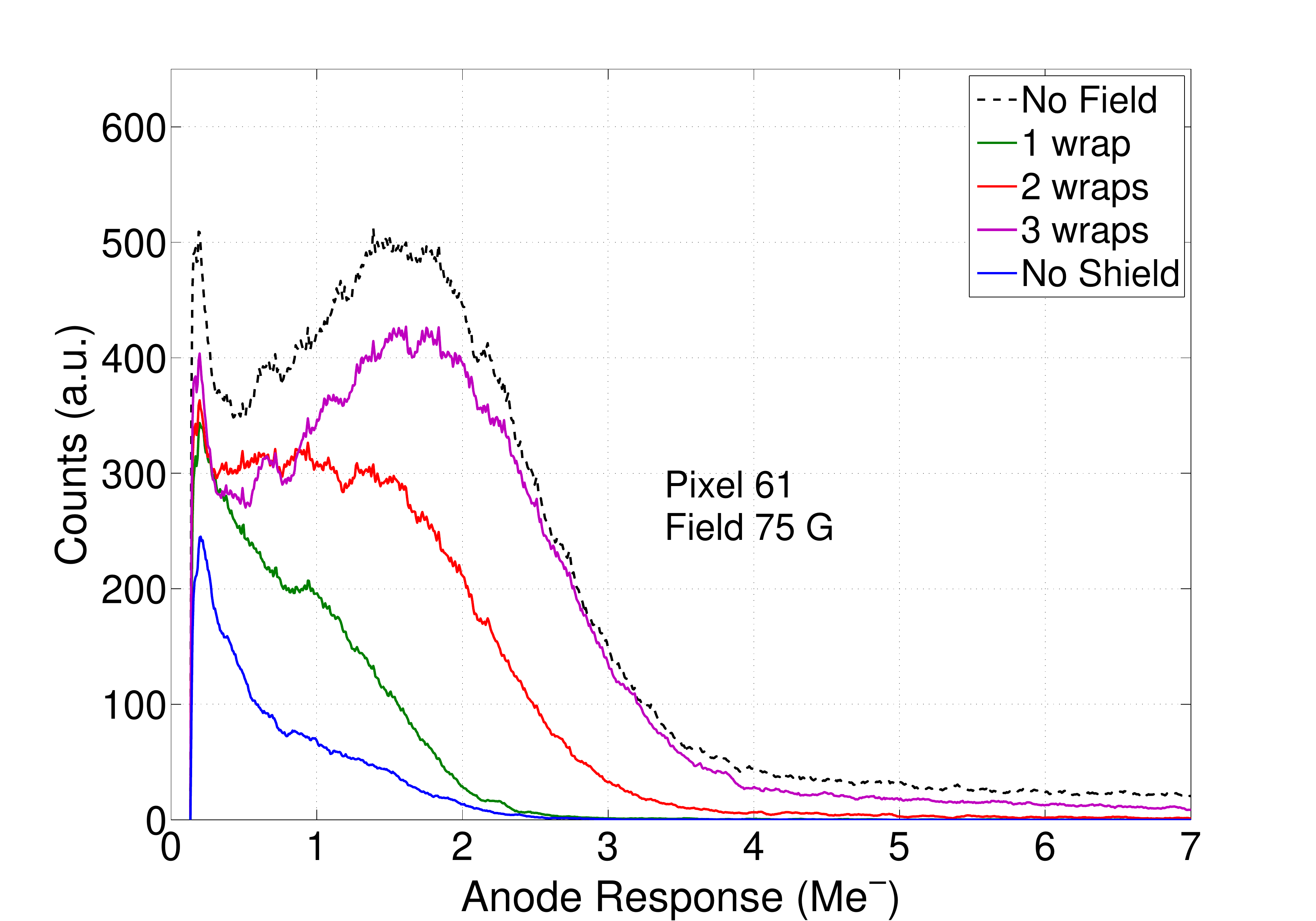}
		\hspace{5mm}%
		\includegraphics[width=0.45\textwidth]{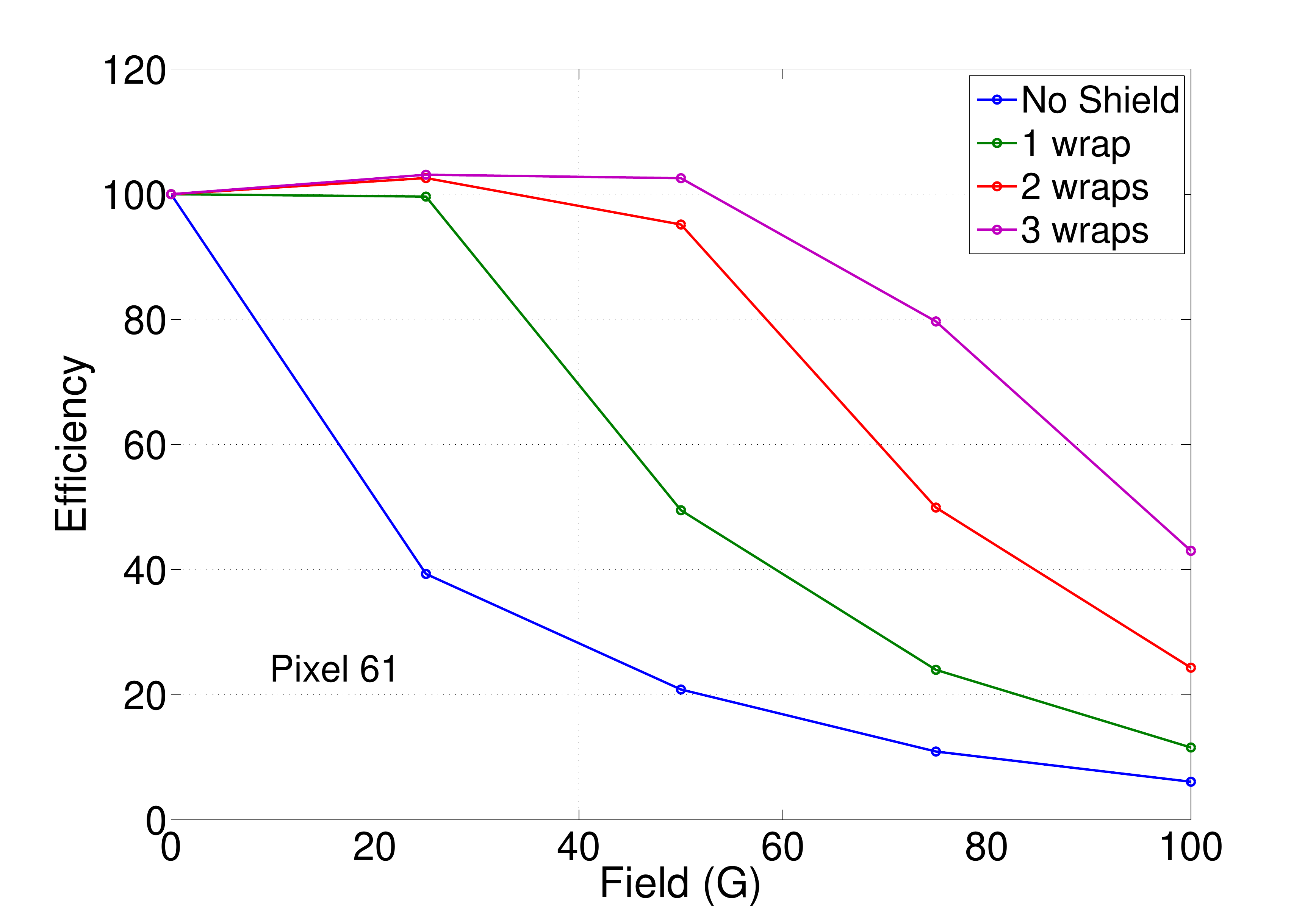}
		\caption{In the first three pictures, the single photon spectra acquired with the MaPMT \itshape R11265-103-M64\upshape (ZN0702) are shown. They are obtained using shields with different thickness. The longitudinal magnetic field ranges from 25 to 75 Gauss. The bottom-right plot shows the photon detection efficiency, defined as the relative rate of events whose amplitude is larger than the noise level ($2\cdot10^5$ $e^-$), as a function of the magnetic field.}
	\label{fig:Thickness}
\end{figure}

Another important conclusion from fig.\ref{fig:ProtrudePixel48} and \ref{fig:ProtrudePixel62} is that, irrespective of the protrusion length, the shield efficiency decreases in case of field larger than 50~G.
This is a typical behaviour of the ferromagnetic material whose relative permeability, as a function of the magnetic field, initially increases, then reaches a maximum value and finally decreases. 
Indeed, over a critical value, field generated within the magnetic shield causes the relative permeability to asymptotically approach unity. 
This behaviour is known as saturation effect and it causes the loss of the shield efficiency at stronger field. 
The critical value over which the shield saturates depend on the material of the shield itself (generally, the higher is the permeability, the weaker is the saturation field), its shape and thickness. 
Hence, in order to obtain good shielding performance at 50~G, we tested thicker shields made of multiple  Skudotech$\textregistered$ wraps. 
Figure \ref{fig:Thickness} shows the results obtained for pixel 61 (an extremely sensitive pixel which represents one of the worst case observed) using single, double and triple Skudotech$\textregistered$ wraps, corresponding to 200 $\mu$m, 400 $\mu$m, 600 $\mu$m thickness. 
In all these cases, the protrusion length was fixed at 1 cm.

One single Skudotech$\textregistered$ wrap properly absorbs a magnetic field up to 25~G and the saturation effect is not reached.
Under the action of 25~G, neither significant spectra distortion nor photon detection efficiency loss are visible. 
At 50~G the single wrap shield is saturated and its efficiency degraded.
The double wraps shield works satisfactorily: the efficiency increases to about 95 \% and the spectra distortions become negligible. 
At stronger field (up to 75~G) also the double wraps shield becomes inadequate and three wraps seem to be necessary.\\

To summarize: the effects of a magnetic field on the performance of a \itshape R11265-103-M64 \upshape MaPMT was thoroughly investigated. 
We studied a longitudinal magnetic field up to 100~G. 
The performance deterioration seems not to be correlated with the bias voltage and the pixel gain. 
While central pixels proved to be almost unaffected, the peripheral pixels turned out to be very sensitive to a longitudinal magnetic field. 
Thus, 200 $\mu$m Skudotech$\textregistered$ shields were tested which wrapped the lateral surface of the device protruding from the photocathode. 
About 1 cm of protrusion and 2 wraps proved to be adeguate to be sure that even the worst pixels works properly up to 50~G. 


\subsection{Aging Test}
It is well-known that long periods of light exposure may deteriorate the MaPMT nominal performance on three different aspects. First, the DC-gain of the tube could decrease due to the wear of the multiplication dynodes. Second, the photocathode efficiency may decrease, and an increasing number of incident photons may not produce any photoelectron, remaining therefore undetected. Finally, the dark current might increase, due to incident radiation which causes the photocathode and dynodes deterioration. The aim of the tests described in the following section is to quantitatively estimate the aging effects of the \itshape R11265-103-M64 \upshape MaPMT. We have fixed the measurement time to 3000 hours\footnote{This is equivalent to two years of LHCb RICH operation considering that the average effective LHCb operation time over 2011 and 2012 amounted to about 1500 hours per year. (https://lhc-statistics.web.cern.ch/LHC-Statistics)}. 

\subsubsection{Setup}
\begin{figure}[ht]
\centering
\includegraphics[width=1\textwidth]{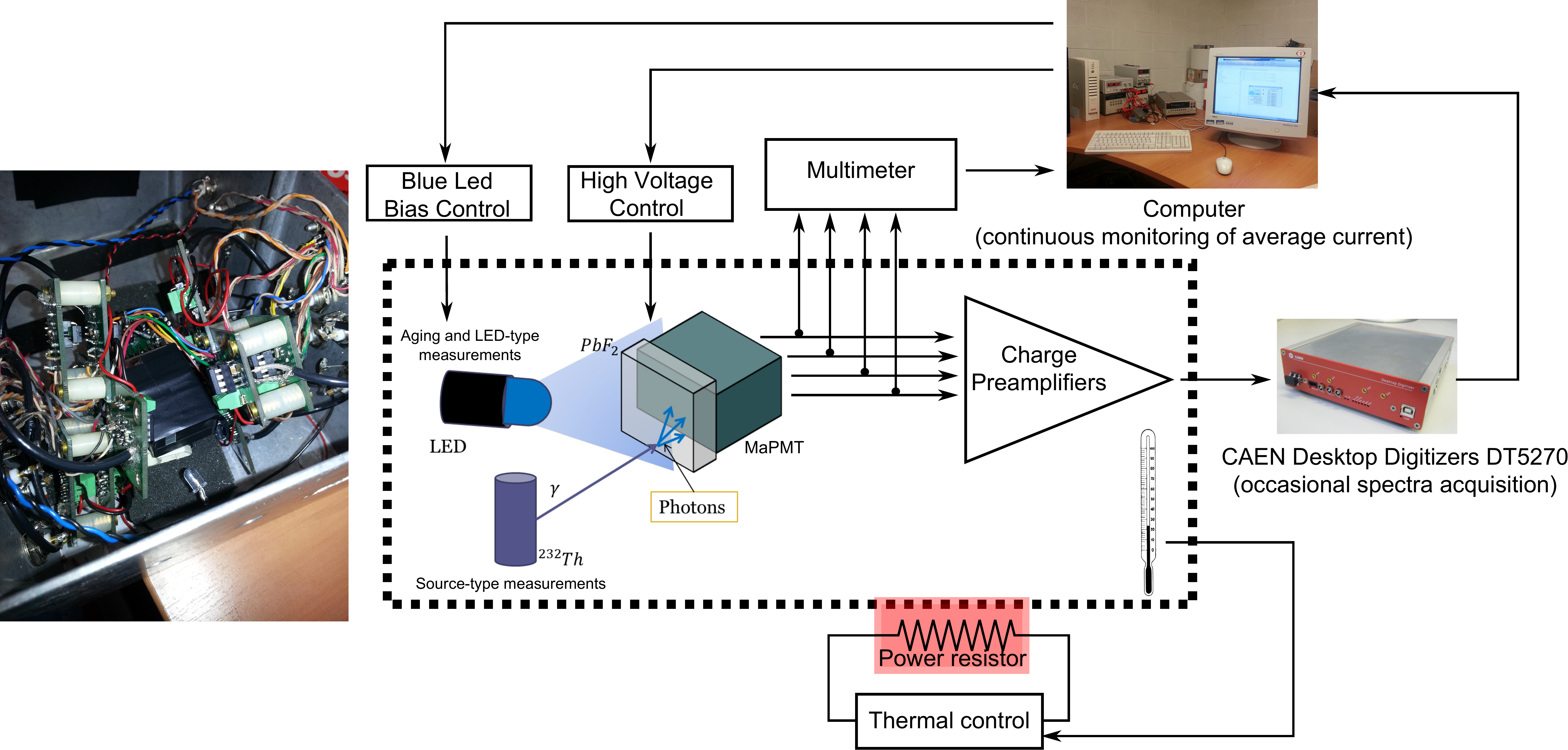}
		\caption{Schematic picture of the fully automatic system developed for the aging test. The MaPMT is covered by a black tape and only a small window leaves the LED light pass. A $PbF_2$ crystal is located in front of the MaPMT and allows to acquire source-type measurements. For each channel, the single photon signal is read out using a charge sensitive preamplifier circuit and acquired by a CAEN Desktop Digitizer. The slow control unit records the DC current and supervises the test ensuring the stability of the high voltage, of the LED supply and of the temperature.}
		\label{Fig:Setup}
\end{figure}

\begin{figure}[h!]
    \centering
		\begin{minipage}[l]{.4825\textwidth}
		\centering
			\includegraphics[width=0.67\textwidth]{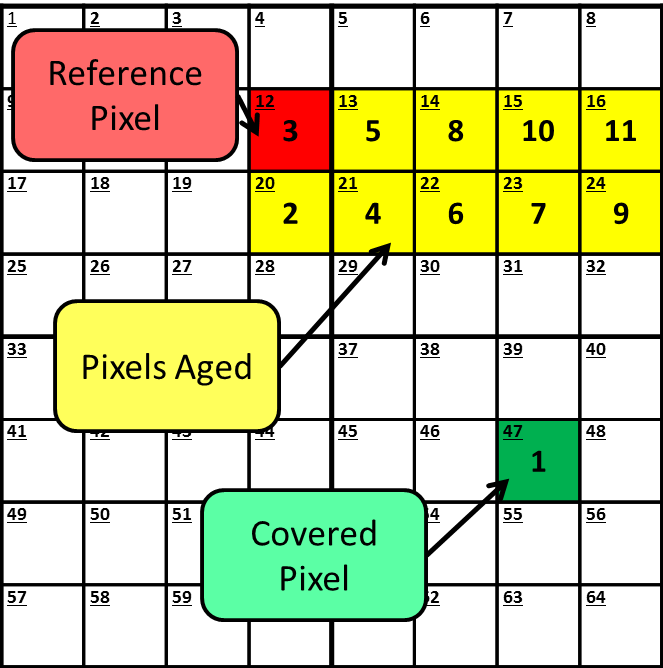}
		\end{minipage}%
		\hspace{15mm}%
		\centering
    \scriptsize
			\begin{tabular}{c||c|c|c}
			 & \multicolumn{3}{c}{MaPMT SN-0707}\\ [0.5ex]
			$CH$&$I_{DC} (\mu A)$&$G (\cdot10^{6})$&$\vartheta (\%)$\\ [0.5ex]
			\hline 
			1 &  0.02 & 1.54 & 0.19\\
			2 &  0.05 & 1.29 & 0.65\\
			3 &  0.18 & 2.08 & 1.35\\
			4 &  0.18 & 1.40 & 2.03\\
			5 &  0.57 & 2.09 & 4.28\\
			6 &  0.57 & 1.26 &	7.05\\
			7 &  0.95 & 1.36 & 10.9\\
			8 &  1.91 & 2.06 & 14.5\\
			9 &  1.63 & 1.33 & 19.1\\
			10 & 2.93 & 2.00 & 22.9\\ 
			11 &  4.91 & 2.02 & 38.0\\
			\hline
			\end{tabular} 
		\caption{Left: Schematic front view of the MaPMT. Right: Table of the studied pixels with the relative DC aging current, gain and occupancy level.}
		\label{fig:pixels}
\end{figure}

In order to observe the aging effects, a long time exposure has to be performed. This requires a setup as automatic as possible able to manage the MaPMT measurements, to check the system stability and to acquire single photon signals. Such system was arranged by developing custom scripts in MATLAB$\textregistered$. Eleven pixels of a \itshape R11265-103-M64 \upshape MaPMT were tested so far and they were illuminated using a commercial blue LED. A $PbF_2$ crystal covered with a black tape was placed in front of the photo sensors, as shown in fig.\ref{Fig:Setup}. A window in correspondence to the read-out pixels was opened on the tape to let the LED light to pass. No optic fiber was used for coupling the LED to the MaPMT so that the pixels were aged with different light exposures (from few tens of nA to about 5 $\mu$A). Figure \ref{fig:pixels} shows the pixels position and a table giving the measured relative average DC aging current, $I_{DC}$, and the expected occupancy percentage $\vartheta$. These values are related by the following equation:
\begin{equation}
\vartheta = \frac{I_{DC}}{R_p \cdot G \cdot q_{e^-}}
\end{equation}
\noindent where $R_p$ is the proton-proton collision rate, $G$ is the pixel gain and $q_{e^-}$ is the electronic charge. 

Considering that the LHC is expected to produce 40 million proton-proton collisions per second ($R_p\simeq40$ MHz) and the gain and the DC aging current were measured for every pixel, the occupancy levels $\vartheta$ can be easily calculated. The measured corrents correspond to an occupancy level of the pixels ranging from 0.1\% to about 38\%.

The system supervised the measurement acquiring the DC aging current values for all pixels every five minutes by a Keithley 2700 Multimeter (fig. \ref{Fig:Setup}). The LED was biased using a Agilent E3631A which was adjusted to keep the aging current stable in one pixel (channel 4) used as the reference. Furthermore, temperature and humidity near the MaPMT were continuously monitored and the temperature was kept stable at 24 degrees using some power resistors located on the black box surface. This configuration prevented any MaPMT gain variation due to temperature fluctuation. 
Finally, the MaPMT bias voltage was periodically measured and fixed at a suitable value which guarantees an adequate gain of about $2\cdot10^6$ electrons per photon.

\subsubsection{Aging measurements}
Every four hours, the system stops the slow control, decreases the LED bias voltage and starts the waveform acquisition. Each observed pixel was connected to a standard charge sensitive preamplifier circuit and the shaped signals, over a fixed threshold of $0.2$~$Me^-$, were acquired by three 4-channel ADCs CAEN Desktop Digitizer DT5720. For each trigger a 1 $\mu s$ waveform was acquired and saved. The off-line analysis calculated the events rate and the signal amplitude as the difference between its maximum and the baseline.
Three different single photon measurements were performed. 
\begin{enumerate}
\item In the LED-type measurement, the LED was biased for about five minutes at a very low voltage so that it operated in a single photon regime compatible with the acquisition rate and the signal to noise ratio. This data allowed to obtain for every pixel the single photon spectrum and to estimate the gain loss from the peak position variation. 
\item The Noise-type measurement was performed turning off the LED for half an hour and acquiring dark signals. 
\item Once per month, a Source-type measurement was performed to qualitatively observe the aging effects on the photocathode efficiency. In this case the LED was turned off for about two hours and a ${}^{232}$Th $\gamma$-source was used to produce Cherenkov photons in the $PbF_2$ crystal placed in front of the MaPMT, acting as a Cherenkov medium. Hence, in the Source-type measurement, the photon production rate is constant, depending only on the source activity. If the photocathode looses efficiency due to aging, i.e. some photons do not produce any photoelectron, then a signal rate reduction would be observed.
\end{enumerate}

\subsubsection{System stability performance}
\begin{figure}[h!]
	\centering
			\includegraphics[width=0.48\textwidth]{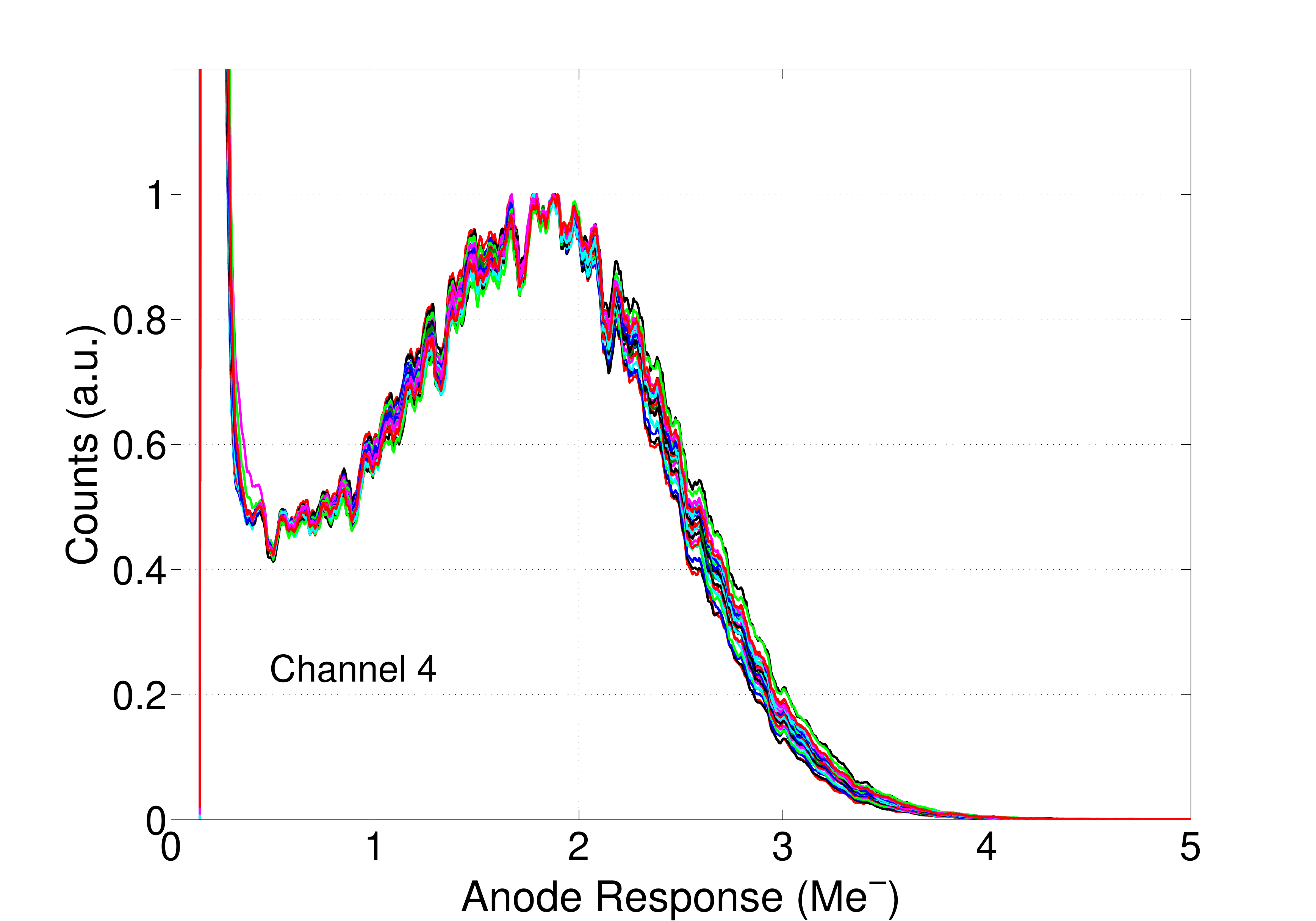}
			\caption{Pre-aging spectra, good superposition.}
			\label{fig:preaging}
	\end{figure}

\begin{figure}[h!]

	\centering
			\includegraphics[width=.47\textwidth]{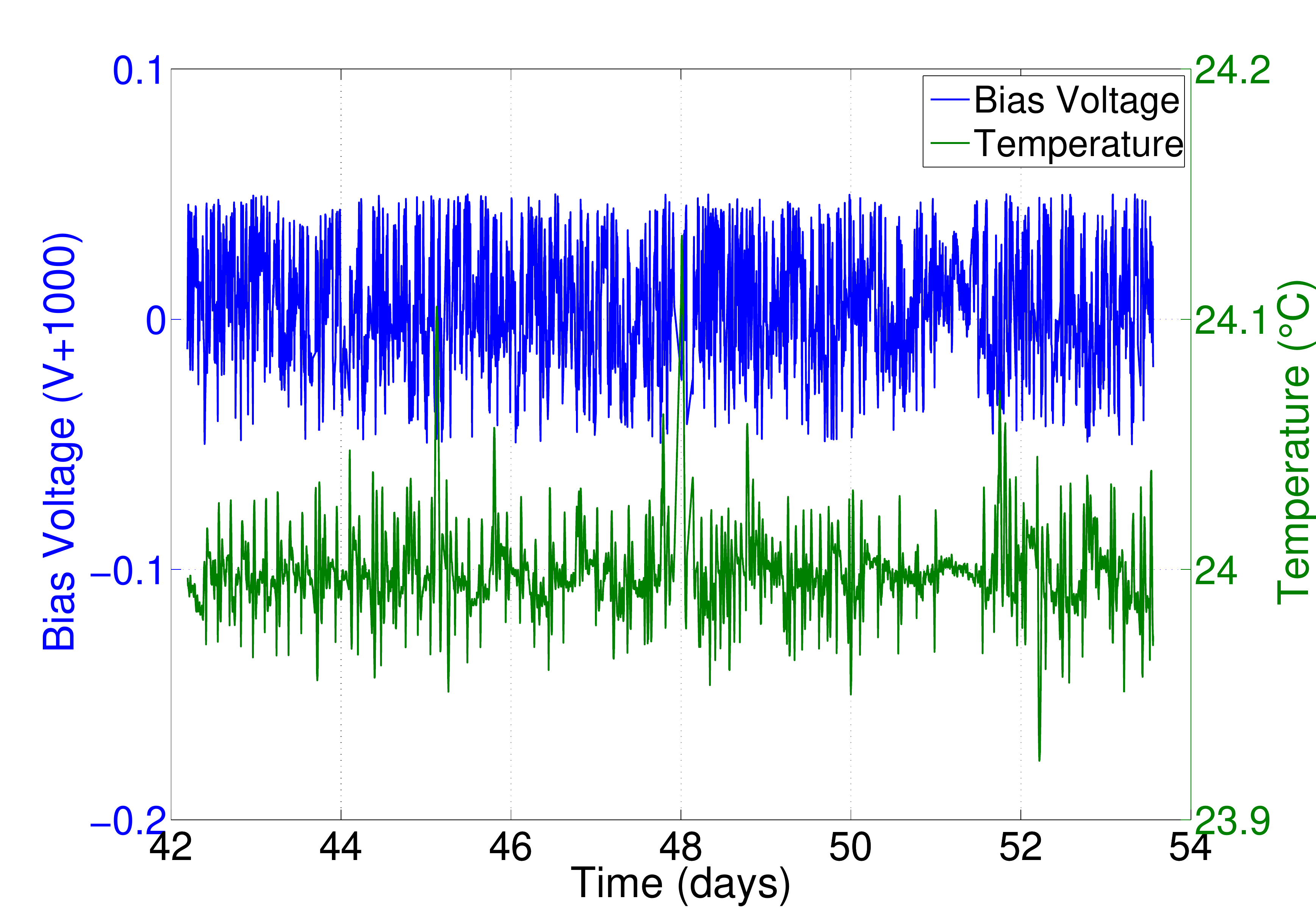}
			\hspace{5mm}%
			\includegraphics[width=.47\textwidth]{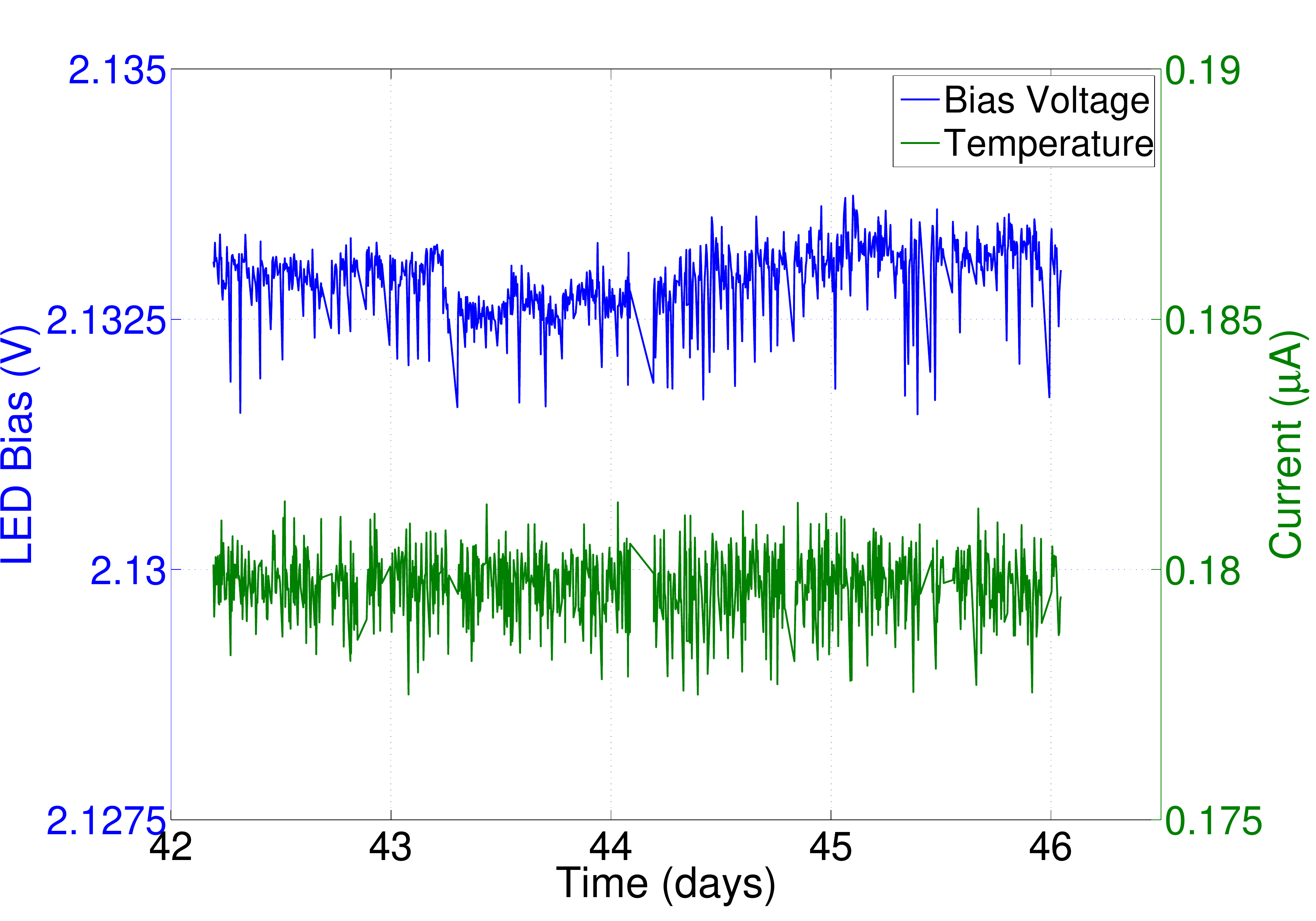}
			\caption{The system stability versus time. Left: high voltage and temperature. Right: LED bias voltage and DC current in the reference channel 4.}
		\label{Fig:Stability}
	\end{figure}

Before starting the aging measurements, the stability of the single photon acquisition chain was checked for three days. Several LED-type measurements (one run every two hours) were performed keeping on the device but without aging it. As expected, all the spectra from each channel were perfectly superimposed (an example is shown in fig.\ref{fig:preaging}). Also few Noise-type measurements were acquired to define precisely the MaPMT starting conditions before its aging.
The stability of the aging process conditions was also continuously monitored. Figure \ref{Fig:Stability} shows the MaPMT bias voltage trend and, the maximum variation observed was 50 mV at -1000 V (0.005\%). The temperature control worked efficiently and ensured a maximum variation of about 0.05 $^\circ C$. Finally, the DC current of the reference pixel (channel 4) versus time is plotted in fig.\ref{Fig:Stability} (right): it was kept stable at 0.180 $\mu A$ with a maximum variation of about 5 nA ($\sim$3\%) adjusting the LED bias voltage.  

\subsubsection{Results}
\begin{figure}[h!]
	\centering
			\includegraphics[width=.49\textwidth]{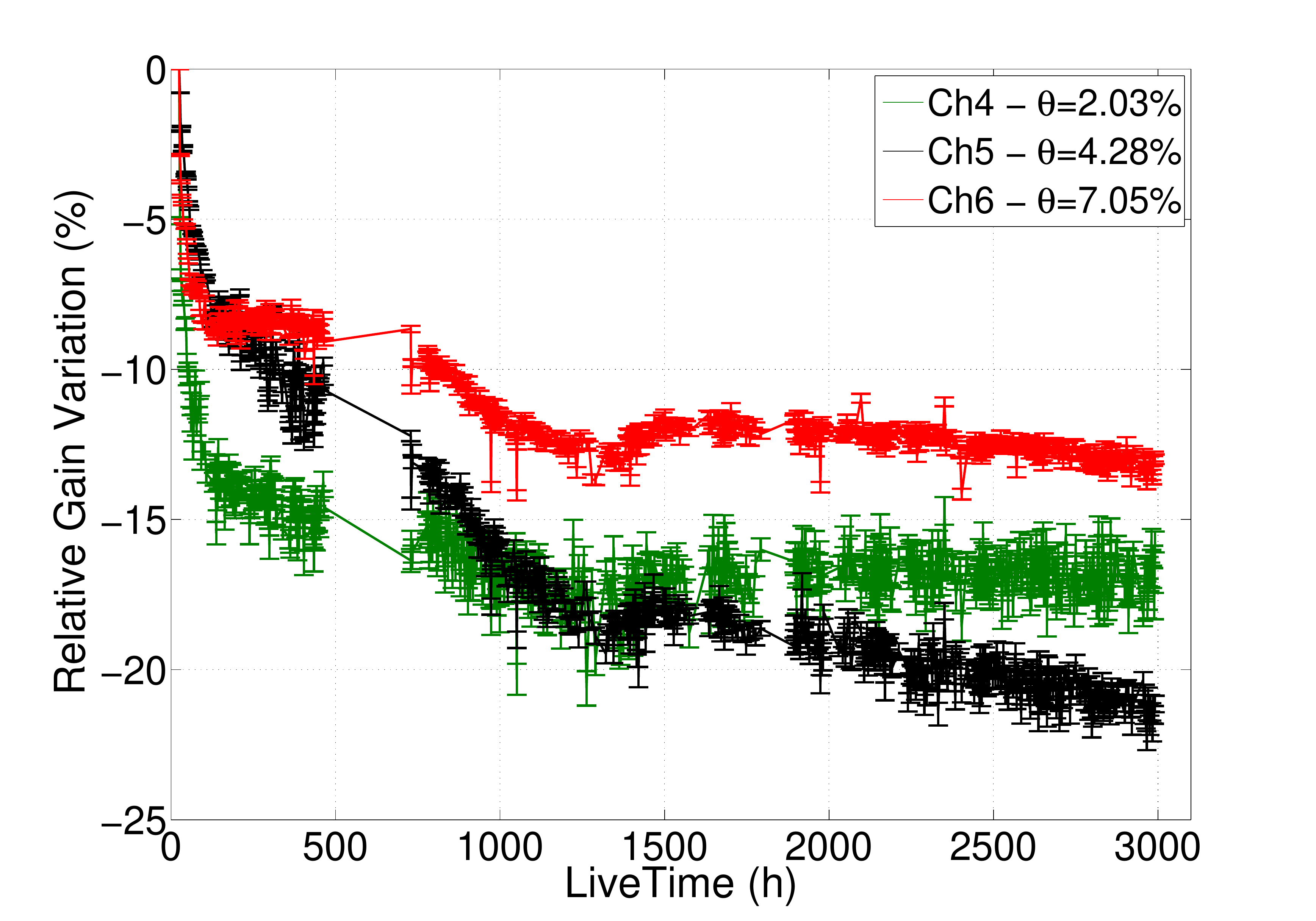}
			\includegraphics[width=.49\textwidth]{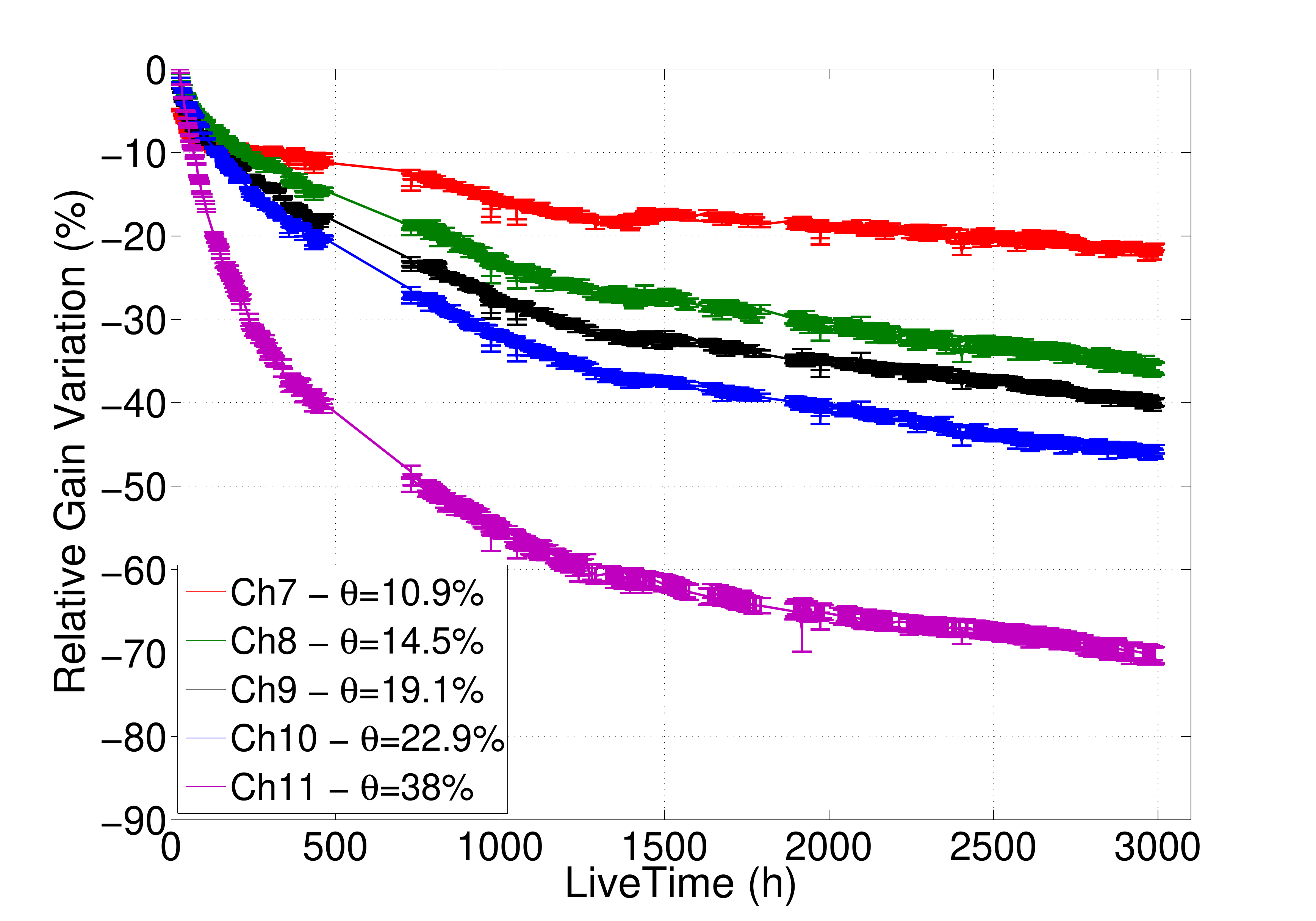}			
	\caption{Gain variation (in \%) versus live-time for some of the tested pixels (SN-ZN0707). Left: The pixels with an occupancy lower than 10\%. Right: The pixels with an occupancy higher than 10\% }
	\label{Fig:Aging}
\end{figure}

Figure \ref{Fig:Aging} shows the results concerning the gain variation due to the aging for most of the observed pixels. Note that the live-time takes into account only the effective period of LED illumination. More than 600 acquisitions were performed over more than 3000 hours. For each acquisition, the gain of each pixel was obtained from the acquired spectra by measuring the energy of the single photon peak. A whole day of illumination turned out to be necessary for stabilizing the MaPMT response, so the percentage gain variation is evaluated considering the gain value at 24 hours as the starting point condition.

\begin{figure}[htbp]
	\centering
		\begin{minipage}[t]{.48\textwidth}
		\centering
		\includegraphics[width=1\textwidth]{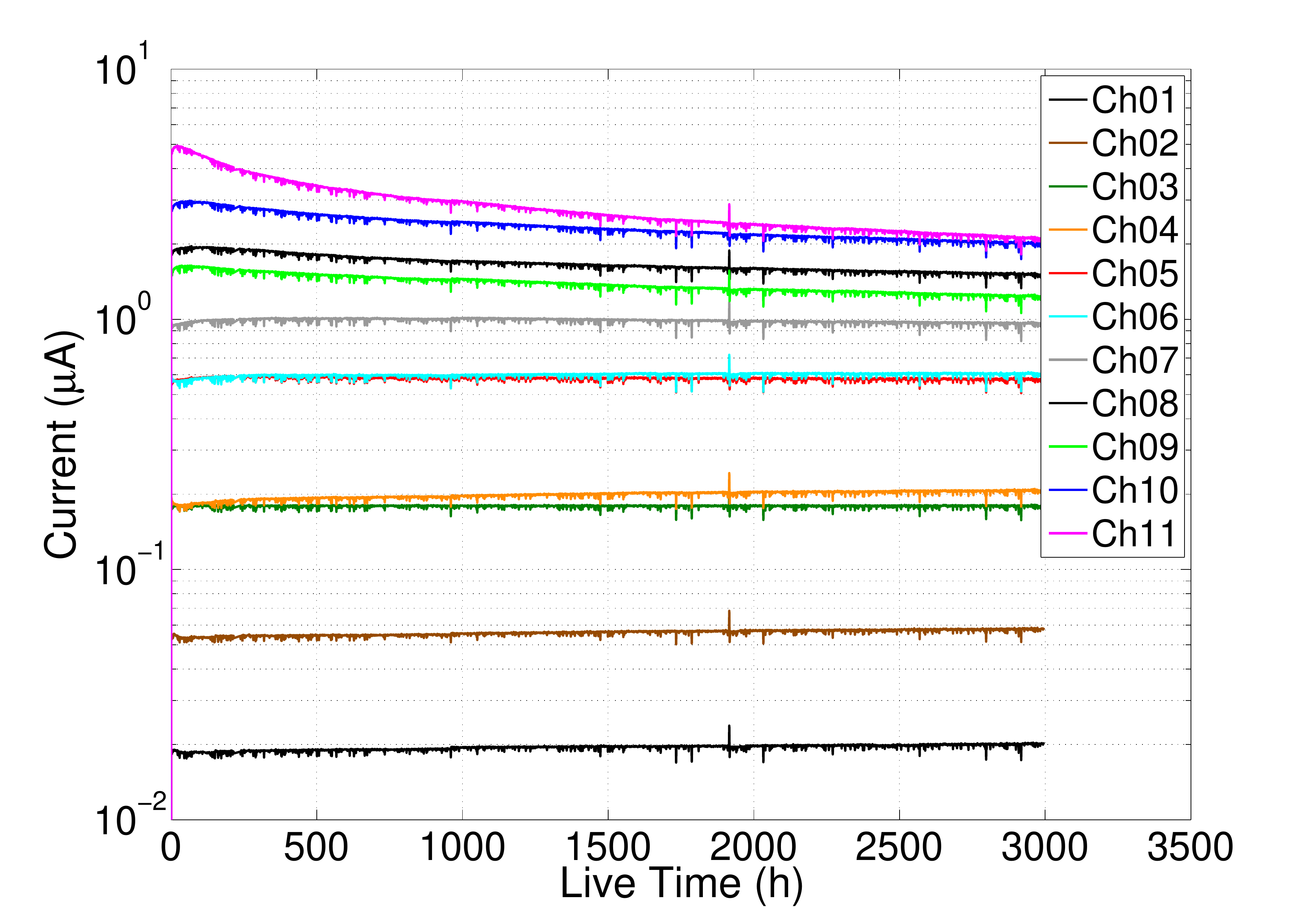}
		\caption{The DC aging current as a function of the live-time for the tested pixels. (SN-ZN0707).}
	\label{fig:Aging2Current}
		\end{minipage}%
	\hspace{5mm}%
		\begin{minipage}[t]{.48\textwidth}
		\centering
			\includegraphics[width=1\textwidth]{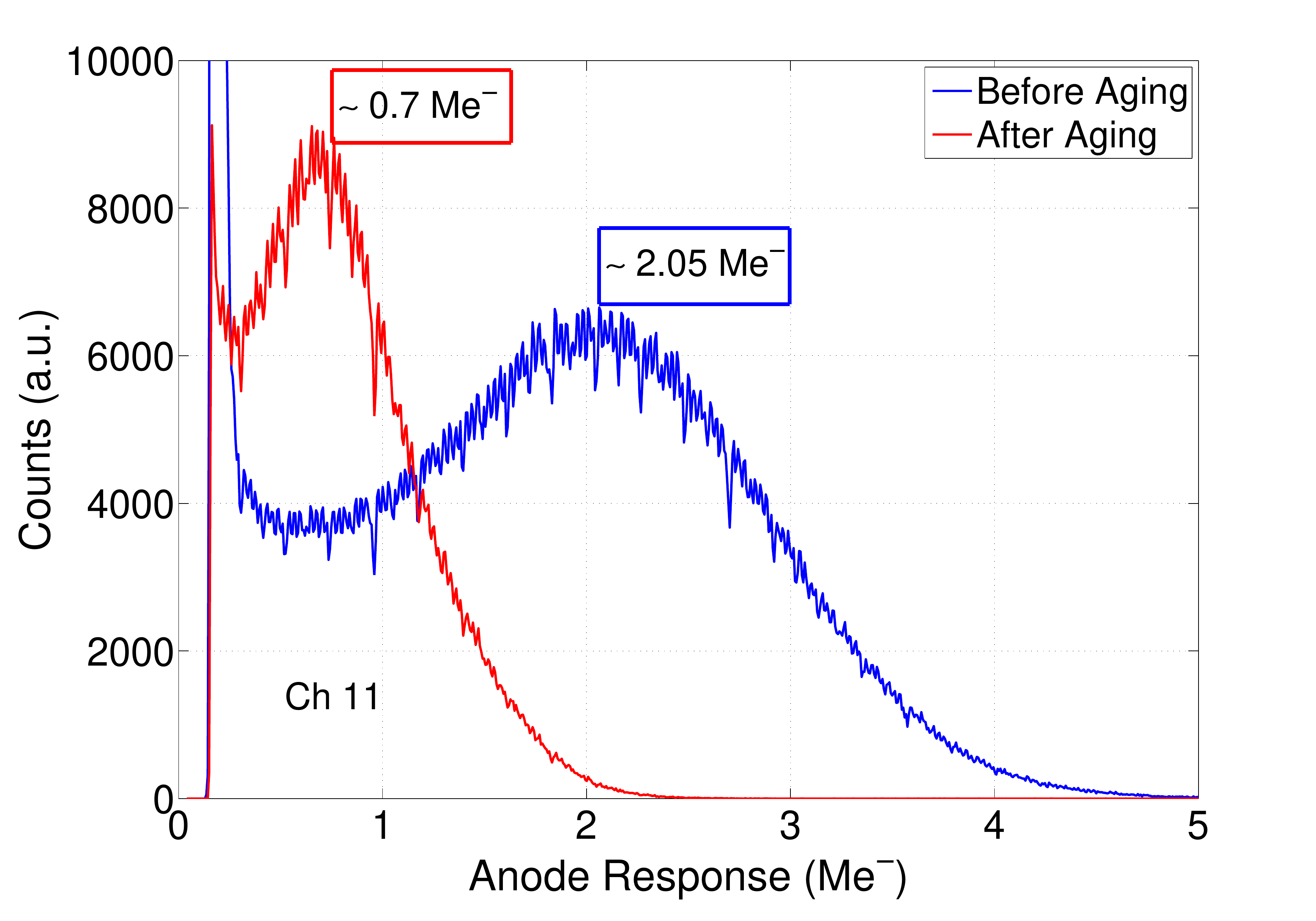}
		\caption{Single photon spectra acquired in the most illuminate pixel (channel 11) before and after the aging.}
			\label{fig:SpectrumAging}
		\end{minipage}
\end{figure}

The gain loss seems to strongly depend on the DC aging current, or equivalent, on the occupancy level $\vartheta$. In particular, the pixels with a DC current lower than 1 $\mu A$ ($\vartheta\lesssim10\%$) have a similar behaviour shown in fig.\ref{Fig:Aging} (left). An initial sharp linear decrease can be observed, followed by a plateau reached after about 1200 hours of LED illumination. The gain loss for these pixels can be assessed at about 12-22\%. 
On the other hand, the gain loss of the most illuminated pixel (channel 11) falls down to about  60\% after only 1200 hours (see fig.\ref{Fig:Aging} right). Roughly, this decrease can be observed also for the other channels with a DC current larger than 1 $\mu A$ ($\vartheta>10\%$). After about 1200 hours, the gain variation reduces, then remaining more stable. A different way of displaying the aging results is to consider the variation of the DC currents measured by the slow control system.  The two methods give comparable results, as fig.\ref{fig:Aging2Current} shows.

\begin{figure}[htbp]
	\centering
		\begin{minipage}[t]{.48\textwidth}
		\centering
			\includegraphics[width=1\textwidth]{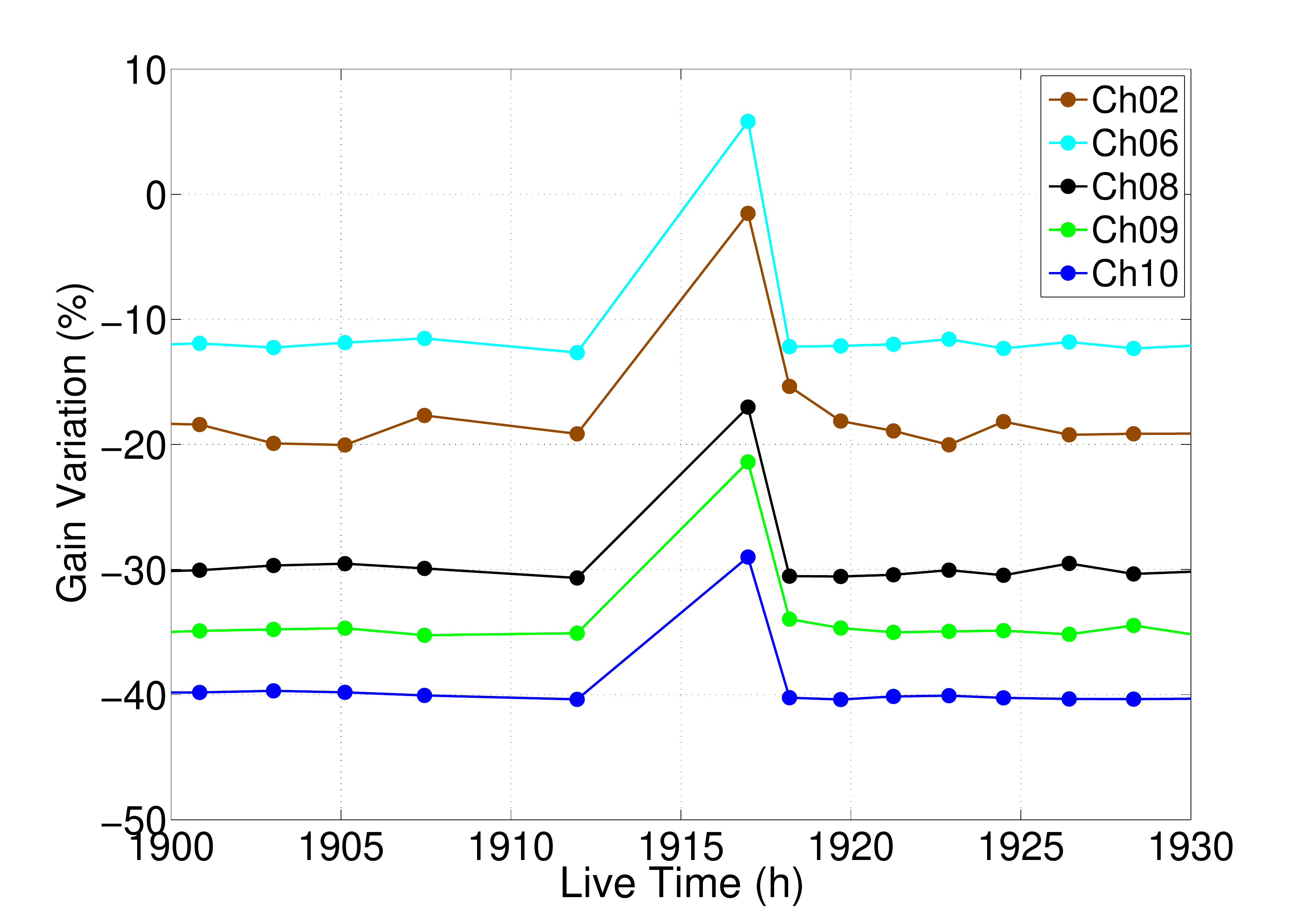}
		\caption{The gain loss decreases by about 10-15\% increasing the bias voltage, from -1000V to -1025V at 1917 hours (SN-ZN0707).}
			\label{fig:Aging2_1025V}
		\end{minipage}%
	\hspace{5mm}%
		\begin{minipage}[t]{.48\textwidth}
		\centering
			\includegraphics[width=1\textwidth]{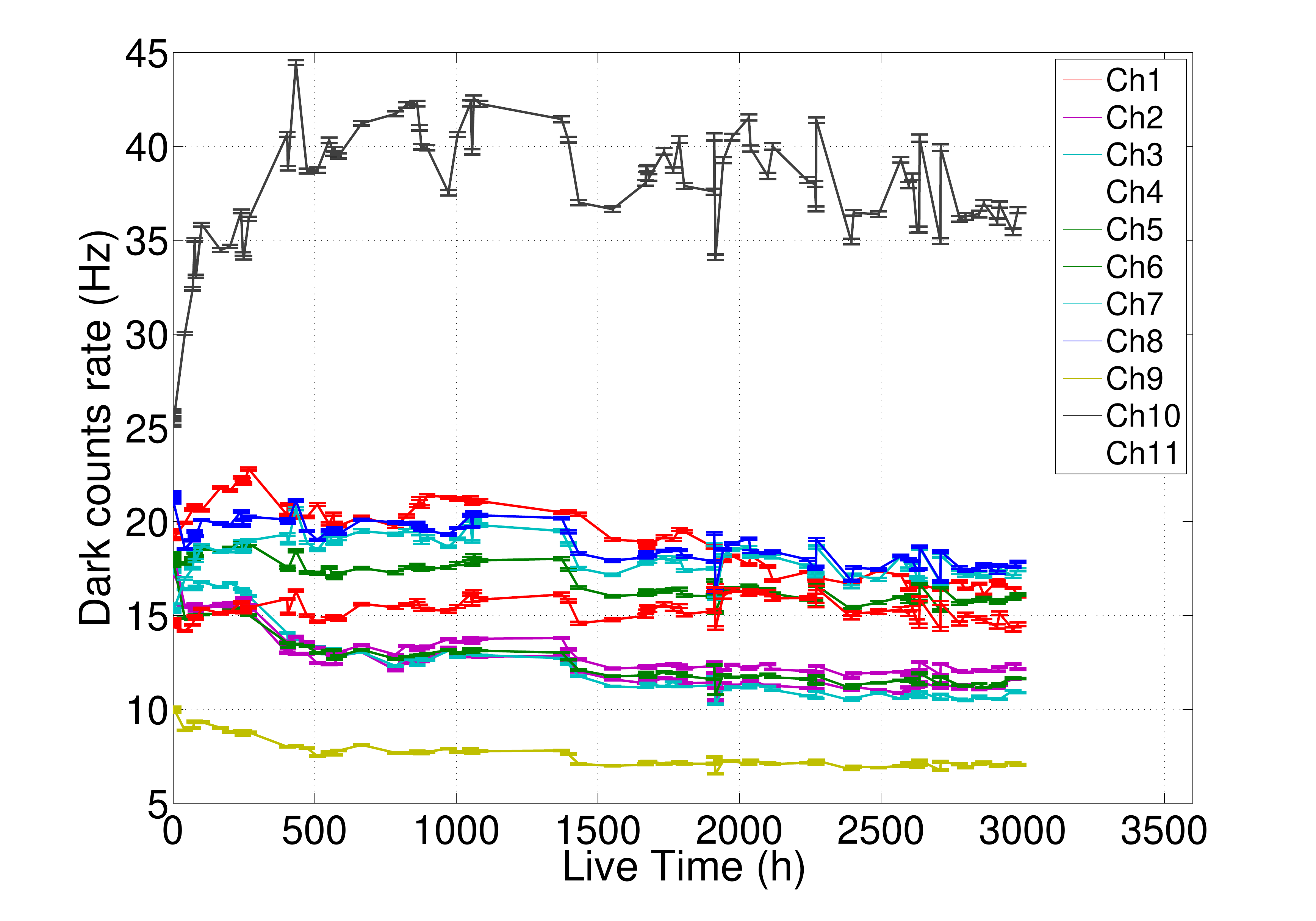}
		\caption{The dark signal events rate for all pixels as a function of live-time.}
			\label{fig:NoiseAging}
		\end{minipage}
\end{figure}

Although the gain loss is higher than expected, even the most illuminated pixel ensures a single photon gain larger than 0.7~Me$^-$ after the aging and the single photon peak is still clearly resolved from the pedestal (fig.\ref{fig:SpectrumAging}). 
The standard method to compensate the aging effect is to increase the bias voltage. As fig.\ref{fig:Aging2_1025V} shows, increasing the high voltage by only 25 V (from -1000 V to -1025 V at 1918 hours) a gain loss of 10-15\% was compensated for almost all the channels. 


Figure \ref{fig:NoiseAging} shows the results of the Noise-type measurements. The initial software threshold over which the events are integrated is fixed as the half of the single photon peak energy, but then it is adjusted taking into account the gain loss obtained by the LED-type measurements. The dark signal rate exhibited a quite similar behavior for all the observed pixels. In particular, the dark signal rate seems not to significantly depend on the aging.

\begin{figure}[h!]
	\centering
			\includegraphics[width=0.48\textwidth]{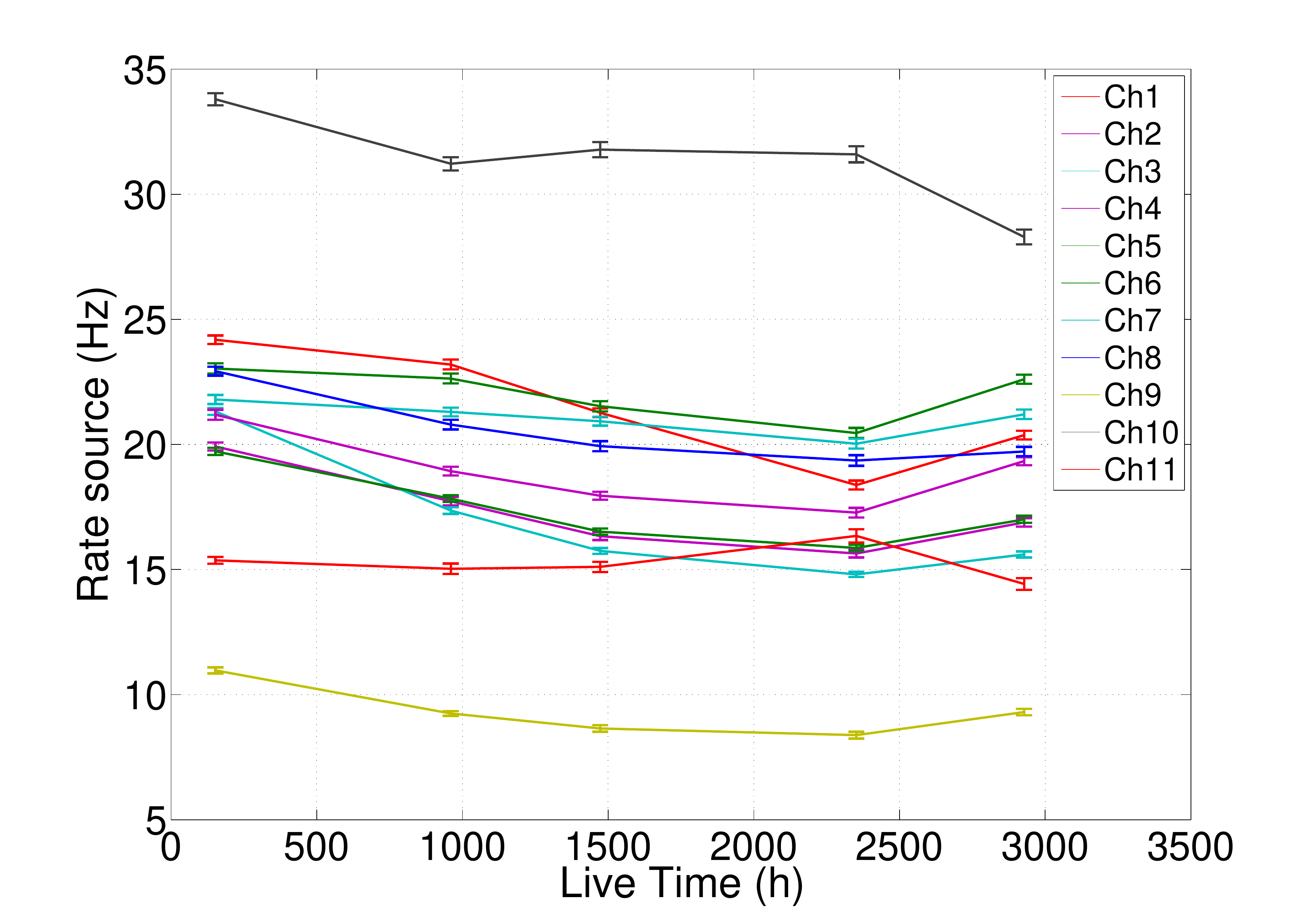}
			\caption{Source event rate as a function of the aging live-time.}
			\label{fig:SourceAging}
	\end{figure}

In the Source-type measurements, the rate of the Cherenkov photon produced in the $PbF_2$ is measured. Similarly to the Noise-type measurements, the trigger threshold is adjusted according to the gain loss. The dark signal contribution is then subtracted in order to estimate the source events rate. Since this setup condition ensures a constant photon production rate of about 20 Hz, if a deterioration of the photocathode efficiency happens, then the observed rate should decrease. Such a measurement does not mean to be a precise study of the photocathode response, but only a qualitatively observed behaviour. As figure \ref{fig:SourceAging} shows, in almost all the pixels the rate decreases by few hertz after $\sim$3000 hours of LED illumination. This suggests that the MaPMT might be affected by a photocathode degradation which would reduce the photon detection efficiency by 5-10\%. Further investigations are necessary for a more precise estimation of this effect.

\section{Conclusions}

We have tested the \itshape R11265-103-M64 \upshape  multi-anode photomultiplier tube produced by Hamamatsu. 
The results on the cross-talk and the dark signal rate have shown that this photomultiplier tube is adequate for an application in high energy physics, such as in RICH detectors. The phototube performance has been studied also as a function of the bias voltage and temperature. A pixel-to-pixel gain spread of about a factor of 3 has been observed, which suggests to couple the phototube with a read-out electronics able to compensate this effect \cite{bib99}. 

A \itshape R11265-103-M64 \upshape MaPMT has been tested with respect to the behaviour in a magnetic field up to 100 G. With a proper high magnetic permeability material shielding, the device can operate well under the action of a longitudinal magnetic field up to 50 G. The basic geometrical properties of the shield have also been investigated. 

The gain variation, the increase of dark signal rate and the loss of the photocathode efficiency due to the aging have been studied. The main aging issue has turned out to be the gain variation, effect which can be compensated by adjusting the bias voltage. 

Further investigations are planned to extensively study the photocathode performance after a long time operation.


\end{document}